\documentclass{aa}  

\usepackage{graphicx}
\usepackage{caption}
\usepackage{subcaption}
\usepackage{xcolor}
\usepackage{placeins}
\usepackage{multicol}

\usepackage{txfonts}
\begin{document}

   \title{TIPSY: Trajectory of Infalling Particles in Streamers around Young stars}

   \subtitle{Dynamical analysis of the streamers around S CrA and HL Tau}

    \author{
    Aashish Gupta\inst{1}\fnmsep\thanks{Aashish.Gupta@eso.org}
    \and
    Anna Miotello\inst{1}
    \and
    Jonathan P. Williams\inst{2}
    \and
    Til Birnstiel\inst{3,4}
    \and
    Michael Kuffmeier\inst{5,6,7}
    \and
    Hsi-Wei Yen\inst{8}
    }
          
    \institute{European Southern Observatory, Karl-Schwarzschild-Str. 2, 85748 Garching bei München, Germany
    \and  Institute for Astronomy, University of Hawaii, Honolulu, HI 96822, USA
    \and  University Observatory, Faculty of Physics, Ludwig-Maximilians-Universität München, Scheinerstr. 1, 81679 Munich, Germany
    \and  Exzellenzcluster ORIGINS, Boltzmannstr. 2, D-85748 Garching, Germany
    \and  Niels Bohr Institute, University of Copenhagen, {\O}ster Voldgade 5, DK-1350 Copenhagen, Denmark
    \and  Department of Astronomy, University of Virginia, Charlottesville, VA 22904, USA
    \and  Max-Planck Institute for Extraterrestrial Physics, Gie{\ss}enbachstra{\ss}e 1, 85748 Garching
    \and  Academia Sinica Institute of Astronomy and Astrophysics, 1, Sec. 4, Roosevelt Rd, Taipei 10617, Taiwan}
    
% \abstract{}{}{}{}{} 
% 5 {} token are mandatory
 
  \abstract
   {Elongated trails of infalling gas, often referred to as "streamers," have recently been observed around young stellar objects (YSOs) at different evolutionary stages. 
   This asymmetric infall of material can significantly alter star and planet formation processes, especially in the more evolved YSOs.
   }
   {In order to ascertain the infalling nature of observed streamer-like structures and then systematically characterize their dynamics, we developed the code TIPSY (Trajectory of Infalling Particles in Streamers around Young stars).}
   { Using TIPSY, the streamer molecular line emission is first isolated from the disk emission. Then the streamer emission, which is effectively a point cloud in three-dimensional (3D) position--position--velocity space, is simplified to a curve-like representation. The observed streamer curve is then compared to the theoretical trajectories of infalling material. 
   The best-fit trajectories are used to constrain streamer features, such as the specific energy, the specific angular momenta, the infall timescale, and the 3D morphology.} 
   % {Using TIPSY, we first isolated the streamer molecular-line emission from the disk emission. Then the streamer emission, which is effectively a point cloud in three-dimensional (3D) position--position--velocity space, was simplified to a curve-like representation. We then compared the observed streamer curve to the theoretical trajectories of infalling material. 
   % The best-fit trajectories were used to constrain streamer features, such as the specific energy, the specific angular momenta, the infall timescale, and the 3D morphology. 
   {We used TIPSY to fit molecular-line ALMA observations of streamers around a Class II binary system, S CrA, and a Class I/II protostar, HL Tau. Our results indicate that both of the streamers are consistent with infalling motion. 
   For the S CrA streamer, we could constrain the dynamical parameters well and find it to be on a bound elliptical trajectory. On the other hand, the fitting uncertainties are substantially higher for the HL Tau streamer, likely due to the smaller spatial scales of the observations. 
   TIPSY results and mass estimates suggest that S CrA and HL Tau are accreting material at a rate of $\gtrsim27$~M$_{jupiter}$~Myr$^{-1}$ and $\gtrsim5$~M$_{jupiter}$~Myr$^{-1}$, respectively, which can significantly increase the mass budget available to form planets.
   }
   {TIPSY can be used to assess whether the morphology and kinematics of observed streamers are consistent with infalling motion and to characterize their dynamics, which is crucial for quantifying their impact on the protostellar systems.}
   % {TIPSY can be used to assess whether elongated streams of material are infalling onto protoplanetary disks, and to characterise their dynamics which is crucial to quantify their impact on the protostellar systems.}

  \keywords{Methods: data analysis, Planets and satellites: formation, Protoplanetary disks, Stars: formation,  ISM: kinematics and dynamics}
  
  \maketitle
%
%-------------------------------------------------------------------

\section{Introduction}  \label{sec:intro}

%Traditional picture of star and planet formation... Symmetrical collapse.. isolated class IIs..
The traditional picture of low-mass star formation assumes that protostars, together with their circumstellar disks, form due to the axisymmetric collapse of dense protostellar cores \citep[e.g.,][]{Shu1977,Terebey1984}. Then, as the surrounding gas envelope disperses, protostars evolve from the embedded Class 0 and I stage to the Class II stage. These Class II systems are then traditionally assumed to evolve in isolation to form planetary systems, such as our Solar System.

However, stars form in turbulent giant molecular clouds, where the 
initial conditions for star and disk formation cannot be represented as isolated non-turbulent spheres \citep[e.g.,][]{Pineda2023, Hacar2023}.
% just add "especially including turbulence"?
Numerical simulations of molecular clouds that follow
the collapse of many protostellar cores show that the star-formation processes can be highly asymmetrical, with material usually falling onto protostellar systems via elongated channels, or "streamers" \citep[e.g.,][]{Padoan2014,Haugbolle2018,Kuznetsova2019,Lebreuilly2021,Pelkonen2021,Kuffmeier2017,Kuffmeier2023}. Recently, with the increased sensitivity of interferometric observations, such streamers have started to be observed around young stellar objects (YSOs) at various evolutionary stages, from the embedded Class 0 and I sources \citep[e.g.,][]{Tobin2012,Yen2014,Tokuda2018,Pineda2020,Thieme2022,Valdivia-Mena2022,Murillo2022,Hsieh2023,Lee2023,Mercimek2023} to the more evolved Class I/II and II sources \citep[e.g.,][]{Tang2012,Akiyama2019,Yen2019,Alves2020,Garufi2022,Huang2020,Huang2021,Huang2022,Huang2023,Gupta2023}. The infalling streamers observed around more evolved YSOs further challenge our assumption that these systems evolve in isolation to form planetary systems; in reality, they are still embedded in large-scale molecular clouds ($\gtrsim1$~pc) and may continue to accrete material from them.

The infall of material in evolved sources can greatly influence the physical and chemical properties of protoplanetary disks and, thus, of the planets they form. For example, the supply of fresh material can help solve the "mass-budget problem" of protoplanetary disks, in which observations suggest that they are typically not massive enough to form the observed planetary systems 
\citep[e.g.,][]{Manara2018,Mulders2021}. 
Moreover, observations \citep{Ginski2021} and simulations \citep{Thies2011,Dullemond2019,Kuffmeier2021} have shown that material falling at these late stages can be dynamically different from the original parental core and can induce misalignments in disks. This can further explain the misalignments observed in evolved planetary systems \citep[e.g.,][]{Albrecht2022}.
Late infall can also bring chemically different material to the system, which can explain the observed chemical diversity among meteorites \citep{Nanne2019N}. 
% Late infall can explain the steep dependence of mass accretion rates in pre-main-sequence stars on their stellar masses \citep{Padoan2005} and infall-induced accretion bursts can resolve the accretion luminosity problem \citep{Padoan2014}. 
% Late-infall, approximated as a Bondi-Hoyle accretion \citep{Bondi1944}, can be responsible for the steep dependence of mass accretion rates in pre-main-sequence stars on their stellar masses \citep{Padoan2005}. 
% Simulations have demonstrated that infall induced disk instabilities can trigger accretion bursts in protostars \citep[e.g.,][]{Vorobyov2005}
% of material can trigger accretion bursts in protostars \citep[e.g.,][]{Vorobyov2005}, which can naturally resolve the accretion luminosity problem \citep[see][]{Kenyon1990}  \citep{Dunham2012,Padoan2014,Jensen2018}  }
Simulations \citep{Vorobyov2005,Dunham2012,Padoan2014,Jensen2018} have shown that infall-induced accretion bursts can naturally resolve the accretion luminosity problem in protostars \citep[see][]{Kenyon1990}.
% Simulations by \citet{Padoan2014} and \citet{Jensen2018} showed that infall-induced accretion bursts can naturally resolve the accretion luminosity problem in protostars \citep[see][]{Kenyon1990}.
% Finally, this phenomena can also explain some of the observed mass accretion rates (Padoan et al. 2005), FU Orionis outbursts (Dullemond et al. 2019), old ($>10$~Myr) disks (Beccari et al. 2015), and disk substructures (Hennebelle et al. 2017; Kuznetsova et al. 2022).
\citet{Kuffmeier2023} demonstrated that infall of material onto Class II systems can also make them seem less evolved, which may affect studies on populations of YSOs.
% Late-infall, approximated as Bondi-Hoyle accretion \citep{Bondi1952}, has also been shown a way to reform disks
Finally, this phenomenon may also produce some of the observed protoplanetary disk substructures, such as rings \citep{Kuznetsova2022}, spirals \citep{Hennebelle2017,Kuffmeier2018}, and vortices \citep{Bae2015}. 

However, the dynamics of observed streamers need to be characterized to asses their impact on the star and planet formation processes. 
This has only been done for a few streamers, using methods such as analyzing velocity gradients along the streamers in position--velocity space \citep[e.g.,][]{Yen2014,Yen2019,Alves2020}, qualitatively comparing infalling trajectories from \citet{Mendoza2009} to the streamer velocity gradients and morphologies \citep[e.g.,][]{Pineda2020,Valdivia-Mena2022,Garufi2022}, and fitting infalling trajectories determined using the Ulrich-Cassen-Moosman (UCM) model \citep[e.g.,][]{Ulrich1976,Cassen1981} to the streamer structures in position--position--velocity (PPV) space \citep{Thieme2022}.
% This has been done only for a few streamers using methods like: analyzing velocity gradients along the streamers in position-velocity (PV) space \citep[e.g.,][]{Yen2014,Yen2019,Alves2020}, qualitatively comparing streamer velocity gradients and morphologies to the infalling trajectories by \citet{Mendoza2009} \citep[e.g.,][]{Pineda2020,Valdivia-Mena2022,Garufi2022}, and fitting infalling trajectories by UCM (Ulrich-Cassen-Moosman) model \citep[e.g.,][]{Ulrich1976,Cassen1981} to the streamer structures in position-position-velocity (PPV) space \citep{Thieme2022}. 
% Although these studies suggest that infalling streamers can transfer significant mass and angular momentum to the protostellar systems, these parameters are still not often well constrained.
These studies suggest that infalling streamers can transfer
significant mass to the protostellar systems. However, these parameters are usually estimated only for embedded sources, and the range of their possible values is generally not well constrained. 
For more evolved Class I/II and II sources, the infalling material can be dynamically unrelated to the protostellar system, and thus we need to explore a wider range of initial configurations to identify the infalling trajectories that best represent observed structures. 
% On the other hand, protostellar masses are better constrained for these less-embedded sources, which are required for these kind of analysis.

% To analyze streamers around evolved YSOs, we ...
To address these issues, we have developed the code Trajectory of Infalling Particles in Streamers around Young Stars (TIPSY)\footnote{https://github.com/AashishGpta/TIPSY}, which was designed to fit theoretical trajectories of infalling gas to molecular-line observations of streamers, without assuming any initial configuration (relative position and velocity) for the gas.
We further used this code to analyze streamers around two evolved sources: the Class II binary system S CrA, for which a $\sim$1000~au streamer-like structure was reported by \cite{Gupta2023}, and the Class I/II system HL Tau, for which a kinematic analysis of a $\sim500$~au streamer \citep{Yen2019} and the corresponding shock observations \citep{Garufi2022} suggest an infalling motion of gas. 
Evolved sources are also more suitable for this kind of analysis because the protostellar masses can be estimated independently of streamer modeling, as done by using spectroscopy for S CrA \citep{Gahm2018} and via the modeling of a Keplerian disk for HL Tau \citep{Yen2019}.
% TIPSY is also suited to analyze streamers around more evolved sources because the protostellar masses can be estimated independently of streamer modelling, as done using spectroscopy for S CrA \citep{Gahm2018} and modelling keplerian disk for HL Tau \citep{Yen2019}.

The fitting methodology employed by TIPSY is detailed in Section \ref{sec:methodology}. Subsequently, we demonstrate TIPSY by using it  to analyze the streamers around S CrA (Sect. \ref{sec:scra}) and HL Tau (Sect. \ref{sec:hltau}). The results are discussed in Sect. \ref{sec:discussion}, and we conclude in Sect. \ref{sec:conclusions}.

\section{Fitting methodology}  \label{sec:methodology}

TIPSY fits theoretical trajectories expected for infalling gas, following the model given in \citet{Mendoza2009}, to the molecular-line observations of streamers. The fitting is done in three-dimensional (3D) PPV space: right ascension (RA), declination (Decl.), and line-of-sight (LOS) velocity or radial velocity (RV); in other words, the morphology and velocity gradient of streamers are fitted simultaneously.
% \LEt{ Verify that your intended meaning has not been changed. ***}
% \LEt{ We do not allow the use of "e.g." or "i.e." within the main text (in parentheses or within figure/table captions is fine). ***}
To define a general initial configuration of infalling gas, we needed to define the 3D position ($\overrightarrow{r_{0}}$) and velocity ($\overrightarrow{v_{0}}$) vectors relative to the protostar (see Fig. \ref{fig:coordinates}). 
Relative position in RA and Decl. direction as well as the relative speed in the LOS direction can be inferred directly from the observations.
% Initial relative position in RA and Decl. direction can be directly inferred from observed streamer morphology. Similarly, initial relative velocity in LOS direction can also be estimated directly .
% Other configuration parameters (speed in LOS direction and seperation in RA and Decl.) can be inferred directly from the observations.
The remaining three parameters, required to define an initial configuration, are the separation in the LOS direction and the relative speeds in the RA and Decl. directions.  
% In general, three free parameters are required to define an initial configuration of infalling gas, relative to the central protostellar system, and these are: separation in LOS direction and speed in RA and Decl. direction.  
For a range of possible initial configurations, we computed theoretical trajectories (see Sect. \ref{sec:model}) and compared them to the observations (see Sect. \ref{sec:code} and Figs. \ref{fig:tipsy_scra} and \ref{fig:tipsy_hltau}). The distribution of free parameters with reasonable fits was used to estimate uncertainties (see Sect. \ref{sec:errors} and Fig. \ref{fig:errors}). Some of the known caveats associated with this kind of analysis are mentioned in Sect. \ref{sec:caveats}.

The idea of comparing streamer observations to the infalling trajectories in the PPV space is similar to the fitting of UCM trajectories to elongated structures around the Class 0 protostar Lupus 3-MMS by \citet{Thieme2022}. However, the UCM model assumes the particle to be in a parabolic orbit with no initial RV \citep[e.g.,][]{Ulrich1976}. Moreover, \citet{Thieme2022} had to make further assumptions about the configuration of infalling gas, for example an initial radius of $10000$~au and a final centrifugal radius of $105$~au. Such assumptions are less likely to be valid for more evolved (Class I/II and II) sources because the infalling material can be dynamically unrelated to the original parental core of the protostellar system. 
% Moreover, \citet{Thieme2022} checked if their was streamer emission close to points on different trajectories whereas we try to fit a curve representative of the observations by 

\subsection{Physical model} \label{sec:model}
One of the first models for gas infalling onto a protostellar system was given by \citet{Bondi1952}; however, it did not consider the rotation of the infalling gas.
Later, the UCM model developed, which provides analytical solutions for the trajectory of a particle infalling around a protostar, assuming that the initial rotation of the particle is about the rotational-axis of the central protostellar system or the "z-axis" \citep{Ulrich1976,Cassen1981,Chevalier1983,Terebey1984,Visser2009,Shariff2022}. This model has been used to analyze infalling motion of material in streamers around young protostellar systems \citep[e.g.,][]{Thieme2022}. 

The boundary conditions used in the UCM model also assume that the infalling gas starts with a zero RV and it is just bound to the protostar, that is to say, it is on a zero energy parabolic orbit. 
However, for any general initial configuration of infalling gas, especially in the context of late infall of material onto a Class II system, these assumptions may not hold true. \citet{Mendoza2009} extended the UCM model to account for possible nonzero initial RVs and energies.
This model has also been used to study kinematics of material in streamers around protostars at different evolutionary stages \citep[e.g.,][]{Pineda2020,Valdivia-Mena2022,Garufi2022}.
The equations derived by \citet{Mendoza2009} to compute positions and velocities of an infalling particle along its trajectory are listed in Appendix \ref{app:equations}

\begin{figure} 
% \centering
\includegraphics[scale=0.31]{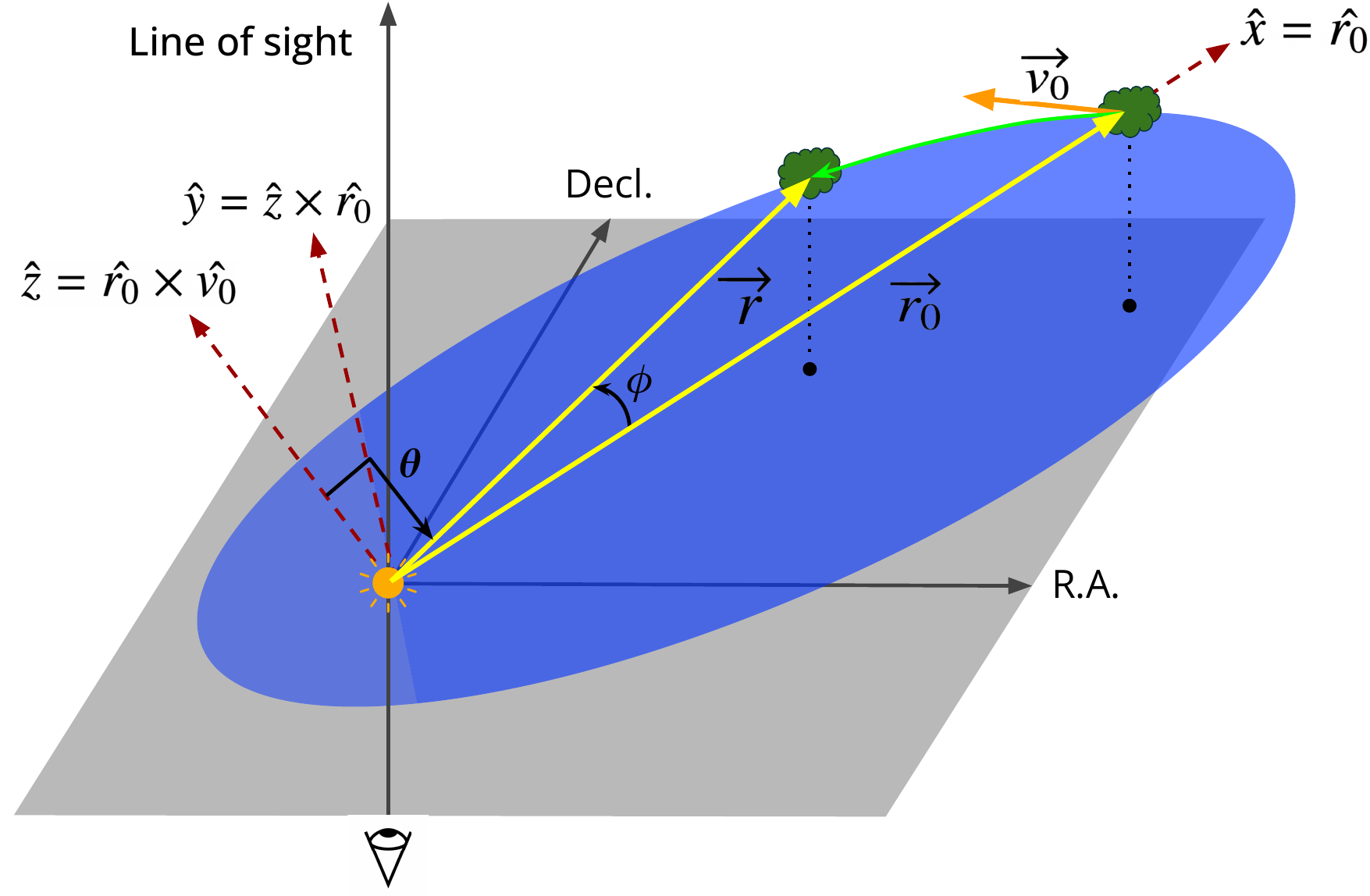}
\caption{Schematic diagram of coordinate axes used to compute the theoretical trajectories of infalling gas (green cloud) around a protostellar system (orange star), as discussed in Sect. \ref{sec:model}. $\overrightarrow{r_{0}}$ and $\overrightarrow{v_{0}}$ denote the initial position and velocity vector of infalling gas, respectively. $\overrightarrow{r}$ represents the position vector of gas at a future point in its trajectory (circumference of blue ellipse), with $\theta$ and $\phi$ denoting the polar and azimuthal angles, respectively. The dashed red arrows show the unit vectors $\hat{x}$, $\hat{y}$, and $\hat{z}$, defined using the directions of $\overrightarrow{r_{0}}$ and $\overrightarrow{v_{0}}$. Together they set the coordinate frame in which TIPSY solves the \citet{Mendoza2009} equations. The gray plane represents the POS, with the overlaid dark gray arrows denoting the coordinate frame of our observations. 
}
\label{fig:coordinates}
\end{figure}

% We use this \citet{Mendoza2009} model to generate infalling traject
However, the original \citet{Mendoza2009} model still assumes the initial rotation is only about the z-axis. This assumption can be mitigated by solving the equations in a rotated coordinate frame, where the z-axis is defined not as the rotational axis of central protostellar system but as a vector normal to the plane of the particle trajectory. This is defined as the plane containing the initial position ($\overrightarrow{r_{0}}$) and velocity vector ($\overrightarrow{v_{0}}$) of the particle with respect to the protostar, as illustrated in Fig. \ref{fig:coordinates}. We used this generalized implementation of the \citet{Mendoza2009} model to generate trajectories of infalling particles without making any assumptions about their initial position or velocity.
The results obtained from our implementation of \citet{Mendoza2009} models were also validated through two-body simulations using the REBOUND framework \citep{rebound}, as shown in Appendix \ref{app:rebound}.

% $\hat{x}=\hat{r_{0}}$\\
% $\hat{y}=\hat{z}\times\hat{r_{0}}$\\
% $\hat{z}=\hat{r_{0}}\times\hat{v_{0}}$
% $\overrightarrow{r_{0}}$
% $\hat{m}$
% $\hat{n}$
% $\hat{r_{0}}$
% $\theta$
% $\phi$
% $\overrightarrow{r}$
% $\overrightarrow{v_{0}}$

\subsection{Fitting procedure} \label{sec:code}

TIPSY is designed to analyze molecular-line observations of streamers with 
% a significant ($\gtrsim3\sigma$) detection of streamer emission over a channel narrow enough (spectral resolution $\sim0.1$~km~s$^{-1}$) to resolve the infalling velocity profile
a large enough recoverable scale to capture the streamer morphology, a sufficient spectral resolution to resolve the velocity profile, and a significant detection ($\gtrsim3\sigma$) of streamer emission in each of the channels 
(see Sect. \ref{sec:requirements} for a further discussion). 
The first step in characterizing streamer observations involves separating the streamer emission from other sources of emission, such as disks. 
% from that of the disk.
Given the wide range of morphologies exhibited by disks and streamers in different sources, which depend on the observational parameters and molecular lines used, it is hard to automate this step. Therefore, we visually examine the emission maps and define a boundary for a sub-cube that encompasses the streamer emission using RA, Decl., and RV limits.
% Given the complexity of different morphologies exhibited by disks and streamers in different sources, observed with different tracers, it is hard to automate this step. Therefore, we visually examine the emission maps and define a boundary for a sub-cube that encompasses the streamer emission using RA, Decl., and RV. limits.
% Because of the wide range of morphologies exhibited by streamers and other sources, which depend on observational parameters and molecular lines used, we examine emission maps visually and define a boundary for a sub-cube that encompasses the streamer emission using RA, Decl., and RV. limits.}
% Due to the wide range of morphologies exhibited by streamer emission and other sources, which depend on observational parameters and molecular lines used, we visually inspect emission maps and define a boundary for a sub-cube that encompasses the streamer emission using RA, Decl., and RV. limits.} 
Next, we eliminate pixels with flux values below a specified noise ($\sigma$) level.
Table \ref{tab:params} lists the values used for selecting streamers around S CrA and HL Tau. 
As long as the selected boundaries fully capture the observed streamer emission, the final results are not very sensitive to the exact values of these limits. This is because we primarily rely on the central brighter emission throughout the structure for the fitting, as described in more detail later.
% a curve-like representation of the observed streamer, based on the central brighter emission throughout the structure, for the fitting, as described in more detail later.

% As we derive a a curve-like representation of the observed streamer structure, the final results are not very sensitive to the details of boundaries selected. In general, it is better to be liberal in defining boundaries  

\begin{table}
% \small
\caption{Parameters used to isolate and fit the HL Tau and S CrA streamers}
\centering
\begin{tabular}{lll}
\hline\hline
Parameter &              S CrA &             HL Tau \\
\hline
Stellar mass [M$_{\odot}$]                         &    2 &    2.1 \\
Distance [pc]                              &    160 &    147 \\
Systemic velocity [km / s]                     &    5.86 &    7.14 \\
Min. RV offset [km / s]                      &    4.5 &    7 \\
Max. RV offset [km / s]                      &    7 &    10 \\
Min. RA offset [arcsec]                      &    2 &    -3 \\
Max. RA offset [arcsec]                      &    15 &    -1 \\
Min. Decl. offset [arcsec]                     &    -7 &    -3 \\
Max. Decl. offset [arcsec]                     &    7 &    0.5 \\
Significance ($\sigma$) level                    &    3 &    4 \\
\hline
\end{tabular} \label{tab:params}
\footnotesize{\\See Appendix \ref{app:stellar_params} for more details on the      stellar parameters used.}
% \caption*{See Appendix \ref{app:stellar_params} for more details on stellar parameters used.}
\end{table}

The resulting sub-cube, comprising mainly of streamer emission, may still contain some unrelated emission features from residual noise or other gas structures.
To get a more cleanly isolated streamer emission, we use a clustering algorithm to identify and remove seemingly unrelated emission. By default, we use the sklearn \citep{scikit-learn} implementation of the  Ordering Points To Identify the Clustering Structure \citep[OPTICS;][]{optics} clustering algorithm, which computes density-based reachability distances to reveal clusters within a dataset. The biggest coherent cluster of emission in the selected sub-cube is identified as the streamer and the smaller clusters generally correspond to noise peaks. 
This step is designed to allow users to set liberal boundaries and noise thresholds while selecting the streamer sub-cube, as noise peaks can then be removed without reducing streamer emission.

\begin{figure*} 
\centering
\includegraphics[width=\textwidth]{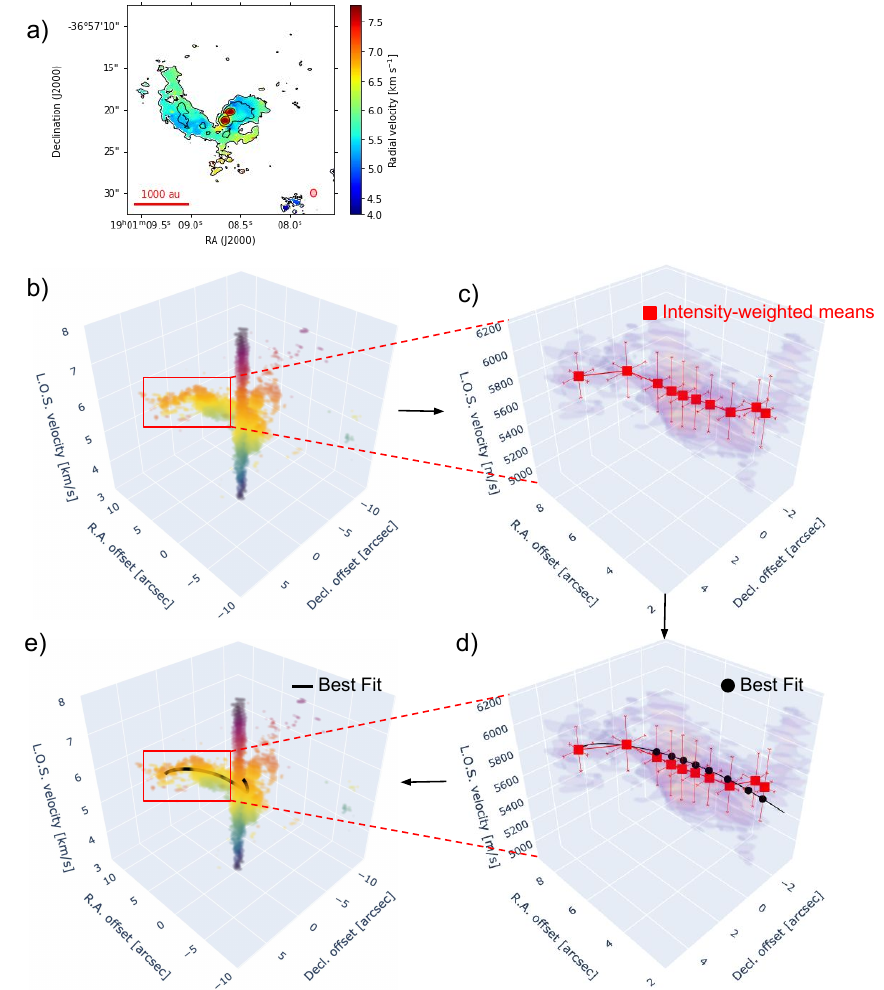}
\caption{Flow of S CrA $^{13}$CO (2--1) data in the TIPSY pipeline.
\textit{Panel a}: Intensity-weighted velocity (moment 1) map in colors, overlaid with contours representing the integrated intensity (moment 0; see Fig. \ref{fig:moment0}).
% Black contours represent integrated intensity (moment 0), starting with $1\sigma$ level and increasing by a factor of 5. 
The red segments in the bottom-left corners depict a length scale of 1,000 au. The pink ellipses in the bottom-right corners depicts the beam size of the data.
\textit{Panel b}: Isometric projection of the 3D PPV diagram of pixels with intensity $>5\sigma$ in the whole field of view.
\textit{Panel c}: Isometric projection of the PPV diagram of an isolated and cleaned streamer. The red square and its error bars represent intensity-weighted means and standard deviations, respectively.
\textit{Panel d}: Same as Panel c, but with the best-fit trajectory, as represented by the black line. Black circles denote the interpolated values of the theoretical trajectory, which are directly compared to the intensity-weighted means.
\textit{Panel e}: Same as Panel b, but with the best-fit trajectory, as represented by the black line.
\textit{Note}: 3D interactive versions of panels d and e are available online.
}
\label{fig:tipsy_scra}
\end{figure*}

\begin{figure*} 
\centering
\includegraphics[width=\textwidth]{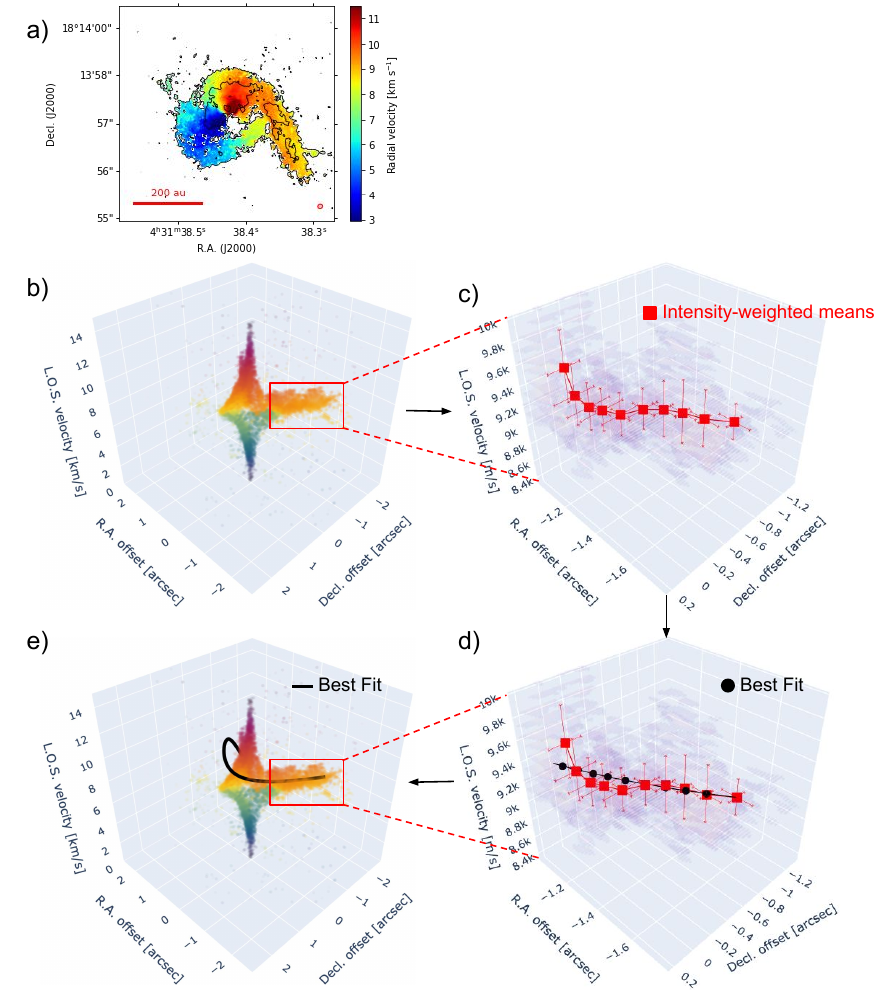}
\caption{
Same procedure as described in Figure \ref{fig:tipsy_scra}, but for HL Tau instead of S CrA.
% Flow of HL Tau HCO$^{+}$ (3--2) data in TIPSY pipeline
% \textit{Panel a}: Intensity-weighted velocity (moment 1) map in colours overlaid with contours representing integrated intensity (moment 0).
% % Black contours represent integrated intensity (moment 0), starting with $1\sigma$ level and increasing by a factor of 5. 
% Red segment in the bottom-left corners depict a length scale of 200 au. Pink ellipse in the bottom-right corners depicts the beam size of the data.
% \textit{Panel b}: Isometric projection of 3D position-position-velocity (PPV) diagram of pixels with intensity $>4\sigma$ in the whole field-of-view.
% \textit{Panel c}: Isometric projection of PPV diagram of isolated and cleaned streamer. Red square and their error-bars represent intensity-weighted means and standard deviations, respectively.
% \textit{Panel d}: Same as Panel c, but with the best fit trajectory as represented by black line. Black circles denote the interpolated values of the theoretical trajectory, which are directly compared to the intensity-weighted means.
% \textit{Panel e}: Same as Panel b, but with the best fit trajectory as represented by black line.
\textit{Note}: 3D interactive versions of panels d and e are available online.
}
\label{fig:tipsy_hltau}
\end{figure*}

This isolated and cleaned streamer emission can be imagined as a point cloud in 3D PPV space, as shown in Figs. \ref{fig:tipsy_scra} and \ref{fig:tipsy_hltau} (panel c). 
The theoretical trajectories of infalling material (as discussed in Sect. \ref{sec:model}) that we aim to fit to the data can be represented as curves in the same 3D space. In order to directly compare the observation to the theoretical curves, we define a curve that would be representative of the observed streamer structure in the PPV space.
To do this, we first divide the streamer points into several bins, set to ten by default, based on a distance metric. 
Then within each of these bins, we compute intensity-weighted means and intensity-weighted standard deviations of the RA, Decl., and RV values of all the points (red squares and their error bars in panel c of  Figs. \ref{fig:tipsy_scra} and \ref{fig:tipsy_hltau}).
This gives us a string of a few points, ten by default, in the same 3D space, which can be directly compared to the theoretical curves (panels c and d in Figs. \ref{fig:tipsy_scra} and \ref{fig:tipsy_hltau}). This method also reduces the dependence of fitting results on the fainter parts of the streamers, selected streamer boundaries, and the spatial and spectral resolution of the data.
As long as an adequate number of bins are used (i.e., enough to capture the overall streamer curvature), the final fitting results will not be sensitive to the number of bins.
% As long as a reasonable number of bins are used, i.e., enough to capture the overall streamer curvature, the final fitting results should not be sensitive on the number of bins used.

% On the other hand, our theoretical trajectories of infalling material, as discussed in Section \ref{sec:model}, can be represented as curves in the same space. In order to better compare them, we simplify the streamer emission to mimic a curve. To do this, we bin the data based on a distance metric (discussed later) and take intensity-weighted means and standard-deviations of all the pixel coordinates with streamer emission, for each bin. This give us a string of a few (ten by default) points in the same 3D space (panel c in Fig. \ref{fig:tipsy_scra} and \ref{fig:tipsy_hltau}), which can be directly compared to the theoretical curves (panel d in Fig. \ref{fig:tipsy_scra} and \ref{fig:tipsy_hltau}).

The distance metric ($d$) we use to bin the data is defined as $d=\sqrt{r^{2}+(wr\theta)^{2}}$, where $r$ and $\theta$ are the polar coordinates of a point on the plane of the sky (POS), with respect to the protostar and the orientation of the streamer very close to the protostar (see Appendix \ref{app:dist} for more details). The $w$ represents a weighting factor to adjust the importance of $r\theta$ distance (azimuthal direction) relative to the $r$ distance (radial direction) in the distance metric calculation and is by default equal to one. 
Overall, larger values of $d$ should denote points in the streamers that are expected to be farther away from the protostar.
Figure \ref{fig:dist} shows the computation of the distance metric values for all the points in the streamers around S CrA and HL Tau.
In addition to the binning of the data, the distance metric is also used as an independent variable for comparing theoretical curves to the observations, as discussed later.

To compare theoretical trajectories with observed streamers, we need to establish a parameter space that covers all the possible initial conditions. For a particle falling onto a protostar, there are seven initial configuration parameters: three for the particle's initial relative position in 3D, three for its initial relative velocity in 3D, and the protostar's mass.
To determine the relative position in the RA and Decl. directions in physical units, we use the physical distance to the protostellar system and the projected separation of the farthest point of the streamer. The separation in the LOS direction is unknown and treated as a free parameter.
For the relative velocity, we use the systemic velocity of the central protostar and the LOS velocity of the farthest point of the streamer to obtain the relative speed in the LOS direction. The relative speed in the RA ($v_{RA}$) and Decl. ($v_{Decl.}$) directions are free parameters. To reduce computations, instead of treating $v_{RA}$ and $v_{Decl.}$ separately, we use the total speed on the POS ($\sqrt{v_{RA}^{2}+v_{Decl.}^{2}}$) and the initial direction of the particle on the POS ($\arctan(v_{Decl.}/v_{RA})$). Here, the initial direction on the POS can be constrained more easily by the projected shape of the streamer.
For evolved sources (Class I/II and II), the protostellar mass is typically assumed to be known from other measurements such as disk rotation \citep[e.g.,][]{Yen2018} or protostellar luminosity \citep[e.g.,][and references therein]{Manara2023}. We note that mass estimates using luminosity can be quite uncertain for young Class I/II sources \citep[e.g.,][]{Baraffe2012}.
In conclusion, we have three free parameters: relative separation in the LOS direction, relative speed on the POS, and the direction of the relative velocity on the POS. TIPSY allows users to set a range of possible values for each of these parameters, which creates a 3D parameter space that is used for the fitting.

Using this parameter space and the \citet{Mendoza2009} model (Sect. \ref{sec:model}), we calculate infalling trajectories for every parameter combinations. These trajectories are compared to the observed streamer curve (intensity-weighted means and standard deviations) to find the best fit.
We independently compare the representative RA, Decl., and RV values using the distance metric, defined earlier as $d=\sqrt{r^{2}+(wr\theta)^{2}}$, as the independent variable for fitting.
We use the first-order spline interpolation, as implemented in scipy \citep{scipy}, to get the theoretical values at the same distance metric values as the points of the observed streamer curve (panel d in Fig. \ref{fig:tipsy_scra} and \ref{fig:tipsy_hltau}).
Then, we examine what fraction of the RA, Decl., and RV values of the observed streamer curve match within the error bars (standard deviations) to the theoretical values.
This fraction is referred to as the "fitting fraction" in Fig. \ref{fig:errors}.
We consider the best-fit trajectory as the one that can accommodate the highest fraction of mean values representing the observed streamer, within their error bars. In cases where multiple trajectories fit the same fraction of values, we choose the trajectory with the lowest chi-squared deviation as the best fit.

\subsection{Error estimation} \label{sec:errors}

\begin{figure*} 
\centering
\begin{subfigure}[b]{1\textwidth}
% \centering
    \includegraphics[width=\textwidth]{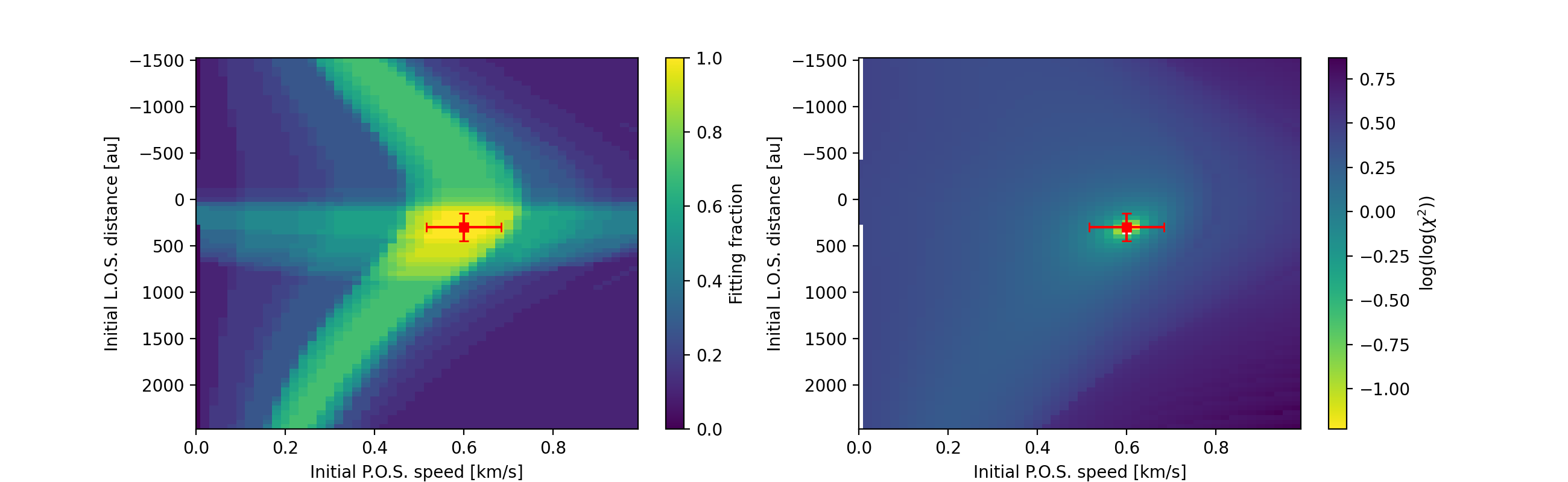}
    \caption{}
    \label{fig:SCrA_errors}
    \end{subfigure}
\hfill
\begin{subfigure}[b]{1\textwidth}
% \centering
    \includegraphics[width=\textwidth]{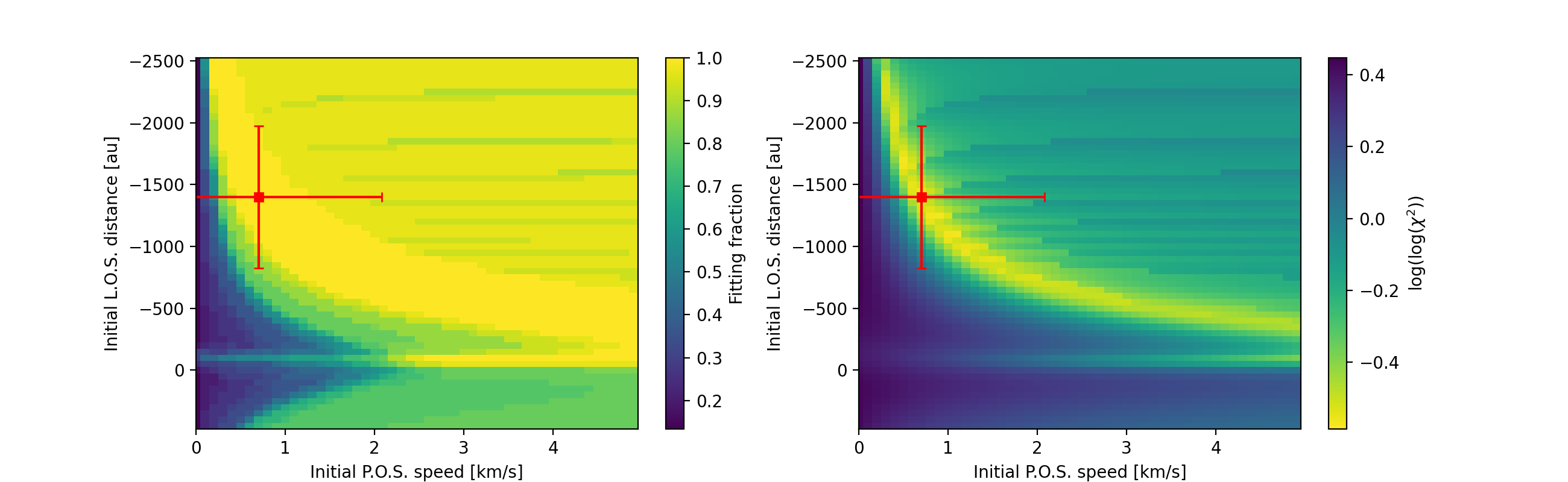}
    \caption{}
    \label{fig:hltau_errors}
    \end{subfigure}
\caption{
Distribution of goodness-of-fit estimates as functions of free parameters: initial speed on the POS (x-axis) and initial spatial offset in the LOS direction, for TIPSY fitting for S CrA (top panels) and HL Tau (bottom panels). Here, initial direction of gas in the POS (third free parameter) is fixed to the value for the best fit.
\textit{Left panels}: Distribution of fractions of coordinate values of points in the observed streamer curve (intensity-weighted means and standard deviations), which is consistent with the theoretical trajectories. 
\textit{Right panels}: Distribution of $\log(\log(\chi^{2}))$ deviations between the observed streamer curve and theoretical trajectories. 
In all the plots, yellow regions represent good fits.
Red squares represent the best fit, as discussed in Sect. \ref{sec:code}. The red lines passing through them represent errors, as discussed in Sect. \ref{sec:errors}.
% Figures showing distribution of goodness-of-fit parameters: fitting fraction (left panels) and $\log(\log(\chi^{2}))$ (right panels) for TIPSY fitting for S CrA (top panels) and HL Tau (bottom panels). Red squares represent the best fit, as discussed in Sec. \ref{sec:code}. Red lines passing them represent errors, as discussed in Sec. \ref{sec:errors}.
}
\label{fig:errors}
\end{figure*}

As discussed in Sect. \ref{sec:code}, we compute theoretical infalling trajectories for each of the parameter combinations and check the fraction of values (RA/Decl./RV) of observed streamer's curve-like representation (intensity-weighted means) that agree with the theoretical values within the error bars (intensity-weighted standard deviations).
To estimate errors in the fitted free parameters (LOS distance, projected speed on the POS, direction on the POS), we select trajectories that can fit at least a certain fraction, 0.9 by default, of the values of observed streamer curve. For each of these trajectories, we store the parameter combinations used to produce them.
Subsequently, the errors are estimated as the standard deviations of each parameter for these parameter combinations with sufficiently good fits (fitting fraction greater $ \geq 0.9$). Figure \ref{fig:errors} display these errors in LOS distances and speed on the POS for the best fits of S CrA and HL Tau.

These error estimates are, by default, also compared to the spatial and velocity resolution of free parameters, used for generating the parameter space. If the error (standard deviation) computed for a parameter is less than the resolution used, the error estimate is increased to this parameter resolution. This generally suggests that the resolution used to create the initial parameter space was too coarse to capture the true fitting uncertainty.

For the parameters estimated directly from the observation (i.e., the offset in RA, Decl. and RV), intensity-weighted standard deviations corresponding to the outermost point of the observed streamer curve is used.
All these uncertainties are further propagated to the derived physical parameters, as listed in Table \ref{tab:results}.
TIPSY also provides a table of goodness-of-fit measurements (fitting fraction and chi-squared deviation) for all parameter combinations, enabling users to independently estimate errors using their preferred methodology.

We note that TIPSY does not currently propagate errors in the fixed parameters to the errors of fitted parameters. These fixed parameters include stellar parameters (stellar mass, systematic velocity, and distance) as well as the parameters corresponding to the observed initial offset of the streamer with respect to to the protostar (offset in RA, Decl., and RV).

% we start with the best fit trajectory parameters, as described in Section \ref{sec:code}. By fixing the other two parameters to the values for the best fit, we vary the parameter of interest and select trajectories that fit at least a certain fraction (default: 0.9) of the observation values. This provides a range of values for the parameter of interest. The standard deviation of these values serves as the error estimate.

% In other words, we assume covariance (contribution of errors in other parameters) is zero while getting errors for each parameter. While this may underestimate uncertainties in general, the covariance between our parameters can be hard to predict as it depends on various streamer properties. Nonetheless, the estimated errors offer insights into the range of initial parameters that can still match the observations. We also provide a table of goodness-of-fit measurements (fraction of fitted values, chi-squared deviation) for all parameter combinations, enabling users to independently estimate errors using their preferred methodology.

\subsection{Caveats} \label{sec:caveats}

An important assumption in the \citet{Mendoza2009} models (see Sect. \ref{sec:model}) used to compute infalling trajectories is that we consider only one force acting on the infalling material: the gravitational force of a point mass (protostar). To begin with, this means that we neglect the contribution of gravitational and tidal 
% \LEt{ We reserve the use of slashes to denote ratios and instrument pairings and for use in equations. The use of "and/or" is acceptable. Kindly rephrase here and where needed. ***}
effects of circumstellar material. 
This assumption is valid as long as most of the mass is concentrated within the central region, which is usually the case, especially in the evolved sources. 
More importantly, we also do not account for the tidal forces from a multiple system, as briefly discussed in Sect. \ref{sec:scra}. We also neglect the effects of gas pressure gradients and shocks \citep[e.g.,][]{Shariff2022}, magnetic fields \citep[e.g.,][]{Unno2022}, and turbulence \citep[e.g.,][]{Seifried2013}. However, these assumptions are generally valid at the length scales of streamers, which are farther away from the protostar, and thus, protostellar systems can be approximated as a point source and the gas density is low. Moreover, the fitting methodology of TIPSY can also be adapted to fit more complicated models to the streamer emission.

% While fitting the observed streamer structures, we also assume that emission from streamer is in optically thin regime. In other words, we assume that the observed intensity of molecular-line emission represents the actual density distribution of gas.
While fitting the observed streamer structures, we also assume that the observed intensity of molecular-line emission represents the actual density distribution of gas. 
This assumption should mostly be valid for low-density streamers; for streamers with a higher density of gas, an appropriately less abundant molecular tracer should be used. 
% The choice of a specific molecular tracer is not necessarily an important assumption for TIPSY, which fits the trajectory of any gas emission detected with an high enough spectral resolution and signal to noise. For what concerns the interpretation of the results, it is on the other hand important to understand what a specific molecule is tracing. 
% In low-density streamers, qualitatively CO is expected to be the tracing the bulk of the gaseous emission, similarly to the case of diffused clouds, and this explains the confusion seen in the 12CO map of S CrA where the clouds also efficiently emit. For this reason, the less abundant isotopologue 13CO is probably a better tracer of the steamer alone. Similarly, to clouds, photodissociation should not be a limiting factor, while freeze-out may be relevant, as temperatures are expected to be quite low (compatible with 20-30 K). The latter is testified by the bright HCO+ emission seen in the HL Tau streamer, as such a molecule becomes abundant when gas-phase CO is depleted, e.g., by freeze out. To provide a more solid and quantitative assessment of the best tracers needed for studying the bulk gas content of a streamer, more complex physical-chemical models should be used, but this is beyond the scope of this work.

Finally, it is important to note that the current implementation of TIPSY does not necessarily rule out the other possible causes of large-scale elongated structures, such as stellar flybys \citep[e.g.,][]{Dong2022,Cuello2023}, the ejection of gas \citep[e.g.,][]{Vorobyov2020}, or gravitational instability in disks \citep[e.g.,][]{Dong2015}.
However, a good fit of observed structures by TIPSY would mean that the observed structures can be explained as infalling streamers. 
Ideally, a comparative analysis with other competing models would be required to identify the most likely cause.
The analytical model used to compute infalling trajectories, as described in Sect. \ref{sec:model}, also allows us to generate unbound hyperbolic trajectories, which may be useful in identifying ejections of unbound gas \citep[e.g.,][]{Vorobyov2020}.
% Some of the competing dynamical processes, like outflow of material may be ruled out,
% based on the velocities which should be greater than the escape velocities (S CrA HL Tau reference...). If the material is outflowing on t
% LOS velocities of outflows can be smaller than escape velocities if they are inclined roughly on the plane-of-sky, which in turn should allow us to use morphology, which should be more ballistic for the outflowing material, to distinguish it from the streamer.
% better distinguish the outflow morphology 
% However, a good fit of observed structures by TIPSY would mean that the observed structures can be explained as infalling streamers. 
Complimentary observations, such as polarization in the near-infrared \citep[e.g.,][]{Ginski2021} and molecules tracing shocks \citep[e.g.,][]{Garufi2022}, can be used to further ascertain the dynamical nature of the streamers.

\section{Applications} \label{sec:applications}

In order to test the TIPSY methodology, we used two protostellar systems with known streamers: the Class II binary source S CrA \citep{Gupta2023} and the Class I/II source HL Tau \citep{Yen2019,Garufi2022}. As TIPSY requires a prior estimation of protostellar mass, it is better suited to analyzing streamers around more evolved Class I/II and II sources. For these sources, most of the streamers have been serendipitously observed in bright $^{12}$CO emission and generally suffer from significant cloud absorption and contamination. This may result in an inaccurate judgement of the extent of the streamers and, thus, an unreliable modeling of them. We present the fitting results for $^{13}$CO (2--1) data of S CrA and HCO$^{+}$ (3--2) data of HL Tau in Sects. \ref{sec:scra} and \ref{sec:hltau}, respectively.

\subsection{S CrA} \label{sec:scra}

S CrA is a binary system, comprising of two Class II sources with a separation of $\sim$ 200~au, in the Corona Australis star-forming region. 
Using spectral and photometric monitoring  of both of the protostars, \citet{Gahm2018} found them to be very similar to each other, with stellar masses of $1$M$_{\odot}$. These masses agree with the modeling of orbital motions by \citet{Zhang2023}. 

\citet{Zhang2023} also reported disk-scale spirals and a $\sim200$~au streamer-like structure connected to the southern protostar, as observed in SPHERE (Spectro-Polarimetric High-contrast Exoplanet REsearch) polarization observations. 
Furthermore, \citet{Gupta2023} found this system to be surrounded by $0.1$~parsec-scale clouds, appearing as reflection nebulae, and $\sim1000$~au elongated structures revealed by $^{12}$CO (2--1) Atacama Large Millimeter/submillimeter Array (ALMA) observations, suggesting infall of material onto the system.
However, these $^{12}$CO observations suffered from contamination from surrounding diffuse gas \citep[see Fig. F.1 in][]{Gupta2023} and also a prominent absorption feature close to systemic velocity of the source, where a lot of bound material is expected to be.

For our analysis we used the $^{13}$CO (2--1) observations, taken as a part of the same ALMA project (Project Id.: 2019.1.01792.S), which show a much cleaner $\sim1300$~au streamer (Fig. \ref{fig:tipsy_scra} and \ref{fig:SCrA_m0}). We used the standard pipeline calibrated data and the imaging was done using "Briggs" weighting with robust=0.5, a cell size of 0.05\arcsec, and "auto-multithresh" masking with default parameters. 
We detected the streamer at $\gtrsim5\sigma$ level, in all the relevant channels.

TIPSY results suggest that the overall streamer is consistent with being a trail of infalling gas. For the best fits, infalling trajectories could fit all the ten points in the simplified streamer (intensity-weighted means, red squares in Fig. \ref{fig:tipsy_scra}) within the error bars (intensity-weighted standard deviations). 
% The best fit parameters, along with the uncertainties, are given in Table \ref{tab:results} (also see Fig. \ref{fig:SCrA_errors}). 
The best-fit parameters, as given in Table \ref{tab:results}, suggest that the material is strongly bound to the protostars, with the specific (per unit mass) total energy (kinetic energy plus gravitational potential energy; see Sect. \ref{sec:physical_params}) of $-1.1\pm0.1$~km$^{2}$~s$^{-2}$. This suggests that the observed structure is not an ejection of unbound material, which should be on hyperbolic trajectories. The size of observed streamer, which is at least an order of magnitude larger than the protoplanetary disks, further indicate that this is not a spiral arm induced by gravitationally unstable disk.

% Our best fit results suggest that the material is strongly bound to the protostars, with the specific (per unit mass) total energy (see Sec. \ref{sec:physical_params}) of $\sim-1.1$~km$^{2}$~s$^{-2}$, suggesting that this material is infalling on the protostellar system.} 
Moreover, we also find that the velocity profile of observed streamer changes close to the protostars, that is to say, the LOS velocities stop decreasing and start increasing, which was not reproduced in our best-fit models (panel d, Fig. \ref{fig:tipsy_scra}). 
This suggests that the gas falling from behind the protostar (see Fig. \ref{fig:SCrA_rebound}) is being slowed.
The change in gas dynamics closer to the protostars could be due to the tidal forces from the binary system that are expected to be dominate in inner regions \citep[e.g.,][]{Zhang2023}.
To test this, a detailed modeling of infalling material interacting with binaries and circumbinary material is required, which is beyond the scope of this study.

\subsection{HL Tau} \label{sec:hltau}

HL Tau is a Class I/II source with a $\sim2$~M$_{\odot}$ mass protostar \citep{Yen2019} surrounded by a protoplanetary disk with concentric rings and gaps \citep{alma2015}. \citet{Yen2019} found the source to be associated with a few-hundred-au-long streamer using HCO$^{+}$ (3--2) ALMA observations. 
They analyzed the velocity gradient along the structure and found it to be dominated by the infalling motion in the outer region. 
HL Tau is also known to be surrounded by a gas envelope with $\sim1000$~au scale asymmetric structures \citep{yen2017}, which may be feeding this streamer.
Furthermore, \citet{Garufi2022} also reported emission from shock tracer (SO$_2$ and SO) at the expected interface of the streamer and the disk, suggesting that the infalling material is impacting the disk.

For our analysis, we used the same self-calibrated HCO$^{+}$ (3--2) ALMA observations (Project Id.: 2016.1.00366.S) of HL Tau as described in \citet{Yen2019}. These observations show significant emission  ($\gtrsim4\sigma$) from the streamer, along with a Keplerian disk, in all the relevant channels (see Figs. \ref{fig:tipsy_hltau} and \ref{fig:hltau_m0}).

TIPSY results, as shown in Fig. \ref{fig:tipsy_hltau}, demonstrate that all the points of the simplified streamer curve (intensity-weighted means) can be fit within the error bars (intensity-weighted standard deviations) by an infalling trajectory. The fitting results are listed in Table \ref{tab:results}. 
For HL Tau, the specific kinetic energy is consistent with the specific gravitational potential energy within the error bars, suggesting that the gas is roughly in a zero-energy parabolic orbit. However, TIPSY could not  constrain the trajectory of infalling particles for HL Tau well (see bottom panels, Fig. \ref{fig:errors}), as further discussed in Sect. \ref{sec:requirements}.

\begin{table}
\small
\caption{Fitting results for S CrA and HL Tau.}
\begin{tabular}{lll}
\hline\hline
Quantity &              S CrA &             HL Tau \\
\hline
RA offset [AU] & 1149$\pm$76 & -227$\pm$12 \\
Decl. offset [AU] & 618$\pm$123 & -159$\pm$15 \\
LOS offset [AU] & 300$\pm$150 & -1400$\pm$573 \\
RA speed [km / s] & 0.37$\pm$0.18 & 0.3$\pm$0.6 \\
Decl. speed [km / s] & -0.47$\pm$0.15 & 0.6$\pm$1.2 \\
LOS speed [km / s] & 0.02$\pm$0.17 & 1.79$\pm$0.20 \\
Specific kinetic energy [km$^{2}$ / s$^{2}$] & 0.18$\pm$0.10 & 1.8$\pm$0.9 \\
Specific potential energy [km$^{2}$ / s$^{2}$] & -1.33$\pm$0.09 & -1.3$\pm$0.5 \\
% Specific total energy [km$^{2}$ / s$^{2}$] & -1.15$\pm$0.13 & 0.5$\pm$1.0 \\
Specific angular momentum [AU km / s] & 791$\pm$218 & 606$\pm$1803 \\
Infall time [yr] & 8301$\pm$1358 & 2724$\pm$1237 \\
\hline
\end{tabular} \label{tab:results}
\end{table}

\section{Discussion} \label{sec:discussion}

\subsection{Data requirements} \label{sec:requirements}

Comparing TIPSY fitting results for S CrA (Sect. \ref{sec:scra}) and HL Tau (Sect. \ref{sec:hltau}), we can see that the uncertainties are much higher for HL Tau (see Table \ref{tab:results}). 
Moreover, Fig. \ref{fig:hltau_errors} shows that the distribution of best-fit parameters (higher fitting fractions, lower $\chi^{2}$ deviations) are not well represented by simple symmetrical errors for HL Tau.
% as they are distributed over different layers.
This is likely because the HL Tau observations, limited by the largest recoverable scale, reveal only a $\sim300$~au part of streamer, much shorter than the $\sim1300$~au streamer around S CrA. This section of streamer is not long enough to capture any curvature in streamer morphology, which helps in constraining the speed of infalling particle on the POS. 
This is useful in breaking the degeneracy between the initial POS speed and the LOS separation and, thus, placing a stringent constraint on the streamer trajectories. 
This suggests that observations with higher recoverable scales ($\gtrsim$1000~au) are better for constraining streamer dynamics. 

The channel width (velocity resolution) for both the S CrA and HL Tau observations is $\sim0.1$~km~s$^{-1}$, which allows TIPSY to resolve the velocity profile \citep[for further discussion, see Appendix D of][]{Gupta2023}.
% \LEt{ Verify that your intended meaning has not been changed. ***}. 
Besides this, TIPSY requires a significant streamer  ($>3\sigma$) emission to be observed in all the relevant channels -- as is the case in the analyzed observations -- in order to distinguish the streamer from the surrounding diffuse gas and the background noise. 

\subsection{Physical parameters} \label{sec:physical_params}

As discussed in Sect. \ref{sec:code}, TIPSY fitting results provide estimates for the initial LOS distance ($d_{LOS}$), the initial projected speed on the POS, and the initial direction on the POS for the infalling gas. The initial speed and direction on the POS can be converted to the initial speed in the RA ($v_{RA}$) and Decl. ($v_{Decl.}$) directions using simple trigonometric relations. These parameters, combined with the initial LOS velocity offset ($v_{LOS}$) and the spatial offset in the RA ($d_{RA}$) and Decl. ($d_{Decl.}$) directions, inferred directly from observations, can provide complete information about the initial configuration of infalling gas relative to the protostar. 

These parameters can be used to derive other physically relevant quantities. For example, specific (per unit mass) kinetic energy can be estimated as $0.5\times(v_{RA}^{2}+v_{Decl.}^{2}+v_{LOS}^{2})$. Similarly, assuming that the local gravitational potential is dominated by the mass of protostellar system, specific gravitational potential energy can be estimated as $-G\times M_{*}/\sqrt{d_{RA}^{2}+d_{Decl.}^{2}+d_{LOS}^{2}}$, where $G$ and $M_{*}$ represent universal gravitational constant and mass of protostellar system, respectively. 
We can sum them to get the specific total energy ($T.E.$), which can tell us if the gas is in a bound elliptical orbit ($T.E.<0$, similar to the streamer around S CrA), a bound parabolic orbit ($T.E.\approx0$, similar to the streamer around HL Tau), or an unbound hyperbolic orbit ($T.E.>0$).
% We note that for S CrA,

Using the initial position ($\overrightarrow{r_{0}}$) and velocity ($\overrightarrow{v_{0}}$) vector of infalling gas, we can also estimate the specific angular momentum as $\overrightarrow{r_{0}}\times\overrightarrow{v_{0}}$.
% Multiplying it by the mass of the streamer, we can get total angular momentum of the infalling material. 
This can be compared to the angular momentum of the disks to quantify the role of infalling material in misaligning the protoplanetary disks, as has been suggested by some hydrodynamic simulations \citep[e.g.,][]{Thies2011,Kuffmeier2021}.
For the S CrA and HL Tau streamers, we find the specific angular momentum magnitudes to be $791\pm218$~AU~km~s$^{-1}$ and $606\pm1803$~AU~km~s$^{-1}$, respectively.
% For S CrA and HL Tau streamers, TIPSY results suggest the specific angular momentum to have a magnitude of $807\pm141$~AU~km~s$^{-1}$ and $476\pm848$~AU~km~s$^{-1}$, respectively.
For reference, the specific angular momentum ($l$) in the outer part of a $100$~au Keplerian disk around a  $2$~M$_{\odot}$ star protostar, similar to HL Tau, should be $\sim421$~AU~km~s$^{-1}$ ($l=\sqrt{GM_{*}R_{d}}$, where $G$, $M_{*}$, and $R_{d}$ are the gravitational constant, the protostellar mass, and the disk radii, respectively).
% For reference, the specific angular momentum ($l$) in the outer part of a Keplerian disk of radius ($R_{d}$) $100$~au around protostar of mass ($M_{*}$) $2$~M$_{\odot}$ star, similar to HL Tau, should be $\sim421$~AU~km~s$^{-1}$ ($l=\sqrt{GM_{*}R_{d}}$, where $G$ is Gravitational constant).

% This quantity can be used to quantify the role of infalling material in misaligning the protoplanetary disks, as has been suggested by hydrodynamic simulations \citep[e.g.,][]{Thies2011,Kuffmeier2021}.

As TIPSY provides the complete trajectory of the infalling gas, until the motion is dominated by the gravitational force, we can also infer the 3D (RA, Decl., and LOS distance) morphology of the infalling streamer, as shown in Fig. \ref{fig:rebound}.
% ... discuss what we see breifly...
These morphologies can further be validated using near-infrared polarization observations, as the degree of polarization in such observations can be correlated to the 3D orientation of dust structures \citep[e.g.,][]{Ginski2021}. A better understanding of the 3D morphology of the streamer can be useful in constraining the location and velocity of impact for material falling onto the disk, which allows us to understand the role of infalling material in creating shocks \citep[e.g.,][]{Garufi2022} and disk substructures \citep[e.g.,][]{Bae2015,Kuznetsova2022}. 
% The trajectories can also be related to the larger-scale ($\sim10,000$~au) gas reservoirs \citep[e.g.,][]{Tang2012, Dullemond2019}.

TIPSY also provides an estimate of the infall timescale for the material, defined as the time taken for the best-fit solutions to reach the point closest to the protostars, starting from the farthest point in the observed streamer. 
We found infall timescales of $8301\pm1358$~yr and $2724\pm1237$~yr for the S CrA and the HL Tau streamer, respectively. This implies that these structures are either short-lived ($\lesssim 10,000$~yr, $<1\%$ of typical disk lifetime) or continuously replenished by larger-scale gas reservoirs. Both S CrA \citep{Gupta2023} and HL Tau \citep{Welch2000} are surrounded by large-scale clouds, which can be feeding these streamers. 
Serendipitously detecting short-lived structures should also be less likely, which could further suggest that these structures survive for longer by accumulating material from surrounding clouds.
Large-scale clouds have been observed around other serendipitously detected streamers \citep[e.g.,][]{Gupta2023}.
We also note that the derived infalling timescales are comparable to the lifetimes of tidal arms induced by stellar flybys \citep[e.g.,][]{Cuello2023}.
% \LEt{ Verify that your intended meaning has not been changed. ***}.

% Moreover, it should be less likely to serendipitously detect short-lived structures.
% of short-lived structures should be less likely.
% Assuming these streamers are isolated, the infall timescales would imply short-lived ($\lesssim 10,000$~yr)
% This would imply that these structures should be short-lived ($\lesssim 10,000$~yr) and therefore, should be hard to detect. Alternatively, these streamers can be continuously replenished by larger-scale gas reservoirs. Both HL Tau

% For S CrA and HL Tau streamers, w
Moreover, the infall timescale can be combined with the mass of the streamer to estimate the mass infall rate. 
A rough lower limit of mass of molecular gas can be estimated from integrated flux ($F_{streamer}$), assuming optically thin emission, as 
\begin{equation}
M_{streamer}\gtrsim\frac{2.37m_{H}4\pi D^{2}F_{streamer}}{A_{trans.}h\nu x_{mol.}f_{u}}    
,\end{equation}
\noindent where $m_{H}$ is the mass of a hydrogen atom, $D$ is the distance to the source, $A_{trans.}$ is the Einstein A coefficient of observed line transition, $\nu$ is the line frequency, $x_{mol.}$ is the abundance of the molecule relative to H$_{2}$, and $f_{u}$ is the fraction of molecules in the upper energy state of the transition \citep[e.g.,][]{Bergin2013}. 
Here, $f_{u}$ can be further computed as $f_{u} = 3e^{E_{u}/T}/Q_{mol.}(T)$, where $E_{u}$ is the upper state energy for the transition, $T$ is the gas temperature, and $Q_{mol.}(T)$ is the partition function for the molecule. 
We took $A_{trans.}$ and $E_{u}$ values for $^{13}$CO (2--1) (S CrA) to be 6.038$\times10^{-7}$~s$^{-1}$ and 15.87~K, and HCO$^{+}$ (3--2) (HL Tau) to be 1.453$\times10^{-3}$~s$^{-1}$ and 25.68~K, respectively, from the Leiden Atomic and Molecular Database \citep{lamda}. The $x_{mol.}$ values were taken to be $1.45\times10^{-6}$ for $^{13}$CO \citep[e.g.,][]{Huang2020} and $10^{-9}$ for HCO$^{+}$ \citep[e.g.,][]{jorgensen04}.
For both sources, we assumed a representative temperature of 25~K, which is typical for gas at these $\sim$100-1000~au scales \citep[e.g.,][]{Jorgensen2005}.
At this temperature, $Q_{mol.}(T=25~K)$ values, interpolated from values provided in Cologne Database for Molecular Spectroscopy \citep{cdms}, were 19.6 and 12.0 for $^{13}$CO and HCO$^{+}$, respectively. We computed 
$F_{streamer}$ to be $1.3\times10^{-20}$~W~m$^{-2}$ for S CrA and $4.2\times10^{-21}$~W~m$^{-2}$ for HL Tau by integrating the flux of the isolated streamer emission (panel c in Figs. \ref{fig:tipsy_scra} and \ref{fig:tipsy_hltau}) over both the position and the velocity.

Using these values, we estimated the streamer masses to be $\gtrsim2.1\times10^{-4}$~M$_{\odot}$ and $\gtrsim1.2\times10^{-5}$~M$_{\odot}$ for S CrA and HL Tau, respectively. 
Although these estimates do not include parts of the streamers beyond the primary beams of the interferometric observations, they can still be used to estimate mass infall rates as $\dot{M}_{inf}=M_{streamer}/T_{inf}$, where $T_{inf}$ refers to the infall time for the observed streamer.
Mass infall rates are found to be 
$\gtrsim2.5\times10^{-8}$~M$_{\odot}$~yr$^{-1}$ (or $\gtrsim27$~M$_{jupiter}$~Myr$^{-1}$) for S CrA and $\gtrsim4.5\times10^{-9}$~M$_{\odot}$~yr$^{-1}$ (or $\gtrsim4.7$~M$_{jupiter}$~Myr$^{-1}$) for HL Tau. 
% These values correspond to mass infall rates ($M_{inf}=M_{streamer}/T_{inf}$) of $\gtrsim1.9\times10^{-8}$~M$_{\odot}$~yr$^{-1}$ (or $\gtrsim20$~M$_{jupiter}$~Myr$^{-1}$) for S CrA and $\gtrsim6.8\times10^{-9}$~M$_{\odot}$~yr$^{-1}$ (or $\gtrsim7$~M$_{jupiter}$~Myr$^{-1}$) for HL Tau. 
Interestingly, these values are comparable to mass accretion rates of pre-main-sequence objects \citep[e.g.,][]{Manara2023}, which have been proposed to be influenced by late-accretion of material from large-scale clouds \citep{Padoan2005}.
% $\sim20$~M$_{jupiter}$~Myr$^{-1}$ for S CrA and $\sim7$~M$_{jupiter}$~Myr$^{-1}$ for HL Tau. 
We note that typical mass accretion rates are generally an order of magnitude higher for Class I sources \citep[e.g.,][]{Enoch2009}, which may be a better comparison for HL Tau.

Moreover, over typical disk lifetimes of a few megayears, these mass infall rates can increase mass available for forming planets by an order of magnitude, which can resolve the apparent mass-budget problem in Class II disks \citep{Manara2018,Mulders2021}.
% These mass rates are high enough to increase the mass of typical disk masses by an order
% This will enable us to investigate whether the mass flow rate from the streamer to the disk is significant enough to account for the apparent lack of mass in Class II disks to form the observed population of planetary systems \citep{Manara2018,Mulders2021}. 
The estimated mass flow rates, along with the chemical characterization of streamers, can also be used to understand their impact in shaping disk chemistry \citep[e.g.,][]{Pineda2020}.
We note that these values should be treated as an order of magnitude estimates. A reliable mass estimation will require modeling multiple molecular-line tracers, which is beyond the scope of this paper.

\section{Conclusions} \label{sec:conclusions}

We have developed a code, TIPSY, to study the gas dynamics in infalling elongated structures, often referred to as streamers. TIPSY is designed to simultaneously fit the morphology and velocity profile of the molecular-line observations of streamers with the expected trajectories of infalling gas.
% \LEt{ Verify that your intended meaning has not been changed. ***} 

To begin with, TIPSY results can be used to judge whether the observations of streamer-like structures are consistent with infalling motion, depending on how well the infalling trajectories fit the streamers.
The dynamical nature of the TIPSY solutions and complementary observations \citep[e.g.,][]{Ginski2021,Garufi2022} can be used to rule out other potential causes, such as stellar flybys \citep[e.g.,][]{Cuello2023}, the ejection of gas \citep[e.g.,][]{Vorobyov2020}, or gravitational instability in disks \citep[e.g.,][]{Dong2015}.
% The infalling nature of the gas can be further ascertained using dynamical nature of best fit trajectories and complementary observations like shocking-tracing molecules 
% Moreover, if the best fit trajectories are all strongly bound elliptical solutions, then ejection of unbound material can be ruled out as 
% If the best fit trajectories are all strongly bound elliptical solutions
% Depending on the energetics of the best fit trajectories and also depending on the energectics of the best fit trajectories..
Then, using the best-fit trajectories, TIPSY provides information about the 3D morphology and kinematics of the infalling gas. This can in turn allow us to estimate parameters such as the infall timescale, the specific angular momentum, the specific total energy, and potentially the expected impact zone of the streamer on the protoplanetary disk. These quantities, combined with a better understanding of overall gas reservoirs, can allow us to study the role of infalling material in replenishing disk masses, impacting disk chemistry, tilting disks, and creating disk substructures.

We tested TIPSY on two objects: a $\sim1300$~au $^{13}$CO streamer around S CrA (a Class II binary system) and a $\sim300$~au HCO$^{+}$ streamer around HL Tau (a Class I/II protostar). 
% Our results suggest that both streamers are consistent with infalling motion.
% , especially given the strongly bound solutions for S CrA and observations of shocks at the interface of streamer and disk for HL Tau. 
For S CrA, we could characterize the dynamics of the streamer well, which seems to be consistent with infalling motion.
% appears to be on a bound elliptical trajectory. 
The negative total energy estimated for the observed streamer, along with the large size compared to the protoplanetary disks, suggests that the observed structure does not represent an ejection of unbound gas or spiral arms induced in the disks.

% the streamer appears to be in a bound elliptical orbit 
% we could well characterise the streamer dynamics and 
% We also see a tentative effect of tidal forces in shaping gas velocities closer to the protostars. 
The streamer around HL Tau is also consistent with infalling motion, which is in agreement with the kinematical analysis by \citet{Yen2019} and the shocks observed by \citet{Garufi2022}. However, the uncertainties estimated on the best-fit parameters are relatively large, indicating that the  observations likely cover a too small spatial scale to provide stringent constraints on the overall trajectory. 
This result is very informative on the type of observations that are needed to study and characterize infalling streamers.

Moreover, S CrA and HL Tau appear to be accreting mass at a rate of $\gtrsim27$~M$_{jupiter}$~Myr$^{-1}$ and $\gtrsim5$~M$_{jupiter}$~Myr$^{-1}$, respectively. If sustained for long enough ($\gtrsim$ 0.1 Myr), such mass infall rates can significantly increase the mass budget available to form planets in evolved sources. 

% Moreover, mass infall rates in both streamers appear to be 
% $\gtrsim10^{-8}$~M$_{\odot}$~yr$^{-1}$ rate, which can significantly increase mass-budget available to form planets in evolved sources. 

% Moreover, mass flow rate in both of the streamers ($\gtrsim10^{-8}$~M$_{\odot}$~yr$^{-1}$) seems to be significant enough to resolve 'mass-budget problem' of .

\begin{acknowledgements}
This work was partly funded by the Deutsche Forschungsgemeinschaft (DFG, German Research Foundation) - 325594231.
H.-W.Y. acknowledges support from National Science and Technology Council (NSTC) in Taiwan through grant NSTC 110-2628-M-001-003-MY3 and from the Academia Sinica Career Development Award (AS-CDA-111-M03).
M.K. is supported by a global postdoctoral fellowship of the H2020 Marie
Sklodowska-Curie Actions (897524) and a Carlsberg Reintegration Fellowship
(CF22-1014).
T.B. acknowledges funding from the European Research Council (ERC) under
the European Union's Horizon 2020 research and innovation programme under grant agreement No 714769 and funding by the Deutsche
Forschungsgemeinschaft (DFG, German Research Foundation) under grants
361140270, 325594231, and Germany's Excellence Strategy - EXC-2094 - 390783311.
%ALMA data
This paper makes use of the following ALMA data: ADS/JAO.ALMA\#2019.1.01792.S and ADS/JAO.ALMA\#2016.1.00366.S. ALMA is a partnership of ESO (representing its member states), NSF (USA) and NINS (Japan), together with NRC (Canada), MOST and ASIAA (Taiwan), and KASI (Republic of Korea), in cooperation with the Republic of Chile. The Joint ALMA Observatory is operated by ESO, AUI/NRAO and NAOJ.

We thank Carlo Manara, Jamie Pineda, Maria Teresa Valdivia-Mena, Amelia Bayo, Lukasz Tychoniec, Karina Mauco Coronado, and Claudia Toci for helpful insights and discussion.  

%Softwares: astropy, trivia, 
This work made use of Astropy\footnote{http://www.astropy.org}: a community-developed core Python package and an ecosystem of tools and resources for astronomy \citep{astropy:2013, astropy:2018, astropy:2022}. The three-dimensional plots were made using TRIVIA\footnote{https://github.com/jaehanbae/trivia} and Plotly\footnote{https://plotly.com/python}.

\end{acknowledgements}

\bibliographystyle{aa} % style aa.bst
\bibliography{refs} % your references refs.bib

\begin{thebibliography}{84}
\expandafter\ifx\csname natexlab\endcsname\relax\def\natexlab#1{#1}\fi

\bibitem[{{Akiyama} {et~al.}(2019){Akiyama}, {Vorobyov}, {Liu}, {Dong}, {de Leon}, {Liu}, \& {Tamura}}]{Akiyama2019}
{Akiyama}, E., {Vorobyov}, E.~I., {Liu}, H.~B., {et~al.} 2019, \aj, 157, 165

\bibitem[{{Albrecht} {et~al.}(2022){Albrecht}, {Dawson}, \& {Winn}}]{Albrecht2022}
{Albrecht}, S.~H., {Dawson}, R.~I., \& {Winn}, J.~N. 2022, \pasp, 134, 082001

\bibitem[{{ALMA Partnership} {et~al.}(2015){ALMA Partnership}, {Brogan}, {P{\'e}rez}, {Hunter}, {Dent}, {Hales}, {Hills}, {Corder}, {Fomalont}, {Vlahakis}, {Asaki}, {Barkats}, {Hirota}, {Hodge}, {Impellizzeri}, {Kneissl}, {Liuzzo}, {Lucas}, {Marcelino}, {Matsushita}, {Nakanishi}, {Phillips}, {Richards}, {Toledo}, {Aladro}, {Broguiere}, {Cortes}, {Cortes}, {Espada}, {Galarza}, {Garcia-Appadoo}, {Guzman-Ramirez}, {Humphreys}, {Jung}, {Kameno}, {Laing}, {Leon}, {Marconi}, {Mignano}, {Nikolic}, {Nyman}, {Radiszcz}, {Remijan}, {Rod{\'o}n}, {Sawada}, {Takahashi}, {Tilanus}, {Vila Vilaro}, {Watson}, {Wiklind}, {Akiyama}, {Chapillon}, {de Gregorio-Monsalvo}, {Di Francesco}, {Gueth}, {Kawamura}, {Lee}, {Nguyen Luong}, {Mangum}, {Pietu}, {Sanhueza}, {Saigo}, {Takakuwa}, {Ubach}, {van Kempen}, {Wootten}, {Castro-Carrizo}, {Francke}, {Gallardo}, {Garcia}, {Gonzalez}, {Hill}, {Kaminski}, {Kurono}, {Liu}, {Lopez}, {Morales}, {Plarre}, {Schieven}, {Testi}, {Videla}, {Villard}, {Andreani}, {Hibbard}, \&
  {Tatematsu}}]{alma2015}
{ALMA Partnership}, {Brogan}, C.~L., {P{\'e}rez}, L.~M., {et~al.} 2015, \apjl, 808, L3

\bibitem[{{Alves} {et~al.}(2020){Alves}, {Cleeves}, {Girart}, {Zhu}, {Franco}, {Zurlo}, \& {Caselli}}]{Alves2020}
{Alves}, F.~O., {Cleeves}, L.~I., {Girart}, J.~M., {et~al.} 2020, \apjl, 904, L6

\bibitem[{Ankerst {et~al.}(1999)Ankerst, Breunig, Kriegel, \& Sander}]{optics}
Ankerst, M., Breunig, M.~M., Kriegel, H.-P., \& Sander, J. 1999, SIGMOD Rec., 28, 49–60

\bibitem[{{Astropy Collaboration} {et~al.}(2022){Astropy Collaboration}, {Price-Whelan}, {Lim}, {Earl}, {Starkman}, {Bradley}, {Shupe}, {Patil}, {Corrales}, {Brasseur}, {N{"o}the}, {Donath}, {Tollerud}, {Morris}, {Ginsburg}, {Vaher}, {Weaver}, {Tocknell}, {Jamieson}, {van Kerkwijk}, {Robitaille}, {Merry}, {Bachetti}, {G{"u}nther}, {Aldcroft}, {Alvarado-Montes}, {Archibald}, {B{'o}di}, {Bapat}, {Barentsen}, {Baz{'a}n}, {Biswas}, {Boquien}, {Burke}, {Cara}, {Cara}, {Conroy}, {Conseil}, {Craig}, {Cross}, {Cruz}, {D'Eugenio}, {Dencheva}, {Devillepoix}, {Dietrich}, {Eigenbrot}, {Erben}, {Ferreira}, {Foreman-Mackey}, {Fox}, {Freij}, {Garg}, {Geda}, {Glattly}, {Gondhalekar}, {Gordon}, {Grant}, {Greenfield}, {Groener}, {Guest}, {Gurovich}, {Handberg}, {Hart}, {Hatfield-Dodds}, {Homeier}, {Hosseinzadeh}, {Jenness}, {Jones}, {Joseph}, {Kalmbach}, {Karamehmetoglu}, {Ka{l}uszy{'n}ski}, {Kelley}, {Kern}, {Kerzendorf}, {Koch}, {Kulumani}, {Lee}, {Ly}, {Ma}, {MacBride}, {Maljaars}, {Muna}, {Murphy}, {Norman}, {O'Steen},
  {Oman}, {Pacifici}, {Pascual}, {Pascual-Granado}, {Patil}, {Perren}, {Pickering}, {Rastogi}, {Roulston}, {Ryan}, {Rykoff}, {Sabater}, {Sakurikar}, {Salgado}, {Sanghi}, {Saunders}, {Savchenko}, {Schwardt}, {Seifert-Eckert}, {Shih}, {Jain}, {Shukla}, {Sick}, {Simpson}, {Singanamalla}, {Singer}, {Singhal}, {Sinha}, {Sip{H{o}}cz}, {Spitler}, {Stansby}, {Streicher}, {{{S}}umak}, {Swinbank}, {Taranu}, {Tewary}, {Tremblay}, {Val-Borro}, {Van Kooten}, {Vasovi{'c}}, {Verma}, {de Miranda Cardoso}, {Williams}, {Wilson}, {Winkel}, {Wood-Vasey}, {Xue}, {Yoachim}, {Zhang}, {Zonca}, \& {Astropy Project Contributors}}]{astropy:2022}
{Astropy Collaboration}, {Price-Whelan}, A.~M., {Lim}, P.~L., {et~al.} 2022, apj, 935, 167

\bibitem[{{Astropy Collaboration} {et~al.}(2018){Astropy Collaboration}, {Price-Whelan}, {Sip{\H{o}}cz}, {G{\"u}nther}, {Lim}, {Crawford}, {Conseil}, {Shupe}, {Craig}, {Dencheva}, {Ginsburg}, {Vand erPlas}, {Bradley}, {P{\'e}rez-Su{\'a}rez}, {de Val-Borro}, {Aldcroft}, {Cruz}, {Robitaille}, {Tollerud}, {Ardelean}, {Babej}, {Bach}, {Bachetti}, {Bakanov}, {Bamford}, {Barentsen}, {Barmby}, {Baumbach}, {Berry}, {Biscani}, {Boquien}, {Bostroem}, {Bouma}, {Brammer}, {Bray}, {Breytenbach}, {Buddelmeijer}, {Burke}, {Calderone}, {Cano Rodr{\'\i}guez}, {Cara}, {Cardoso}, {Cheedella}, {Copin}, {Corrales}, {Crichton}, {D'Avella}, {Deil}, {Depagne}, {Dietrich}, {Donath}, {Droettboom}, {Earl}, {Erben}, {Fabbro}, {Ferreira}, {Finethy}, {Fox}, {Garrison}, {Gibbons}, {Goldstein}, {Gommers}, {Greco}, {Greenfield}, {Groener}, {Grollier}, {Hagen}, {Hirst}, {Homeier}, {Horton}, {Hosseinzadeh}, {Hu}, {Hunkeler}, {Ivezi{\'c}}, {Jain}, {Jenness}, {Kanarek}, {Kendrew}, {Kern}, {Kerzendorf}, {Khvalko}, {King}, {Kirkby}, {Kulkarni},
  {Kumar}, {Lee}, {Lenz}, {Littlefair}, {Ma}, {Macleod}, {Mastropietro}, {McCully}, {Montagnac}, {Morris}, {Mueller}, {Mumford}, {Muna}, {Murphy}, {Nelson}, {Nguyen}, {Ninan}, {N{\"o}the}, {Ogaz}, {Oh}, {Parejko}, {Parley}, {Pascual}, {Patil}, {Patil}, {Plunkett}, {Prochaska}, {Rastogi}, {Reddy Janga}, {Sabater}, {Sakurikar}, {Seifert}, {Sherbert}, {Sherwood-Taylor}, {Shih}, {Sick}, {Silbiger}, {Singanamalla}, {Singer}, {Sladen}, {Sooley}, {Sornarajah}, {Streicher}, {Teuben}, {Thomas}, {Tremblay}, {Turner}, {Terr{\'o}n}, {van Kerkwijk}, {de la Vega}, {Watkins}, {Weaver}, {Whitmore}, {Woillez}, {Zabalza}, \& {Astropy Contributors}}]{astropy:2018}
{Astropy Collaboration}, {Price-Whelan}, A.~M., {Sip{\H{o}}cz}, B.~M., {et~al.} 2018, \aj, 156, 123

\bibitem[{{Astropy Collaboration} {et~al.}(2013){Astropy Collaboration}, {Robitaille}, {Tollerud}, {Greenfield}, {Droettboom}, {Bray}, {Aldcroft}, {Davis}, {Ginsburg}, {Price-Whelan}, {Kerzendorf}, {Conley}, {Crighton}, {Barbary}, {Muna}, {Ferguson}, {Grollier}, {Parikh}, {Nair}, {Unther}, {Deil}, {Woillez}, {Conseil}, {Kramer}, {Turner}, {Singer}, {Fox}, {Weaver}, {Zabalza}, {Edwards}, {Azalee Bostroem}, {Burke}, {Casey}, {Crawford}, {Dencheva}, {Ely}, {Jenness}, {Labrie}, {Lim}, {Pierfederici}, {Pontzen}, {Ptak}, {Refsdal}, {Servillat}, \& {Streicher}}]{astropy:2013}
{Astropy Collaboration}, {Robitaille}, T.~P., {Tollerud}, E.~J., {et~al.} 2013, \aap, 558, A33

\bibitem[{{Bae} {et~al.}(2015){Bae}, {Hartmann}, \& {Zhu}}]{Bae2015}
{Bae}, J., {Hartmann}, L., \& {Zhu}, Z. 2015, \apj, 805, 15

\bibitem[{{Baraffe} {et~al.}(2012){Baraffe}, {Vorobyov}, \& {Chabrier}}]{Baraffe2012}
{Baraffe}, I., {Vorobyov}, E., \& {Chabrier}, G. 2012, \apj, 756, 118

\bibitem[{{Bergin} {et~al.}(2013){Bergin}, {Cleeves}, {Gorti}, {Zhang}, {Blake}, {Green}, {Andrews}, {Evans}, {Henning}, {{\"O}berg}, {Pontoppidan}, {Qi}, {Salyk}, \& {van Dishoeck}}]{Bergin2013}
{Bergin}, E.~A., {Cleeves}, L.~I., {Gorti}, U., {et~al.} 2013, \nat, 493, 644

\bibitem[{{Bondi}(1952)}]{Bondi1952}
{Bondi}, H. 1952, \mnras, 112, 195

\bibitem[{{Cassen} \& {Moosman}(1981)}]{Cassen1981}
{Cassen}, P. \& {Moosman}, A. 1981, \icarus, 48, 353

\bibitem[{{Chevalier}(1983)}]{Chevalier1983}
{Chevalier}, R.~A. 1983, \apj, 268, 753

\bibitem[{{Cuello} {et~al.}(2023){Cuello}, {M{\'e}nard}, \& {Price}}]{Cuello2023}
{Cuello}, N., {M{\'e}nard}, F., \& {Price}, D.~J. 2023, European Physical Journal Plus, 138, 11

\bibitem[{{Dong} {et~al.}(2015){Dong}, {Hall}, {Rice}, \& {Chiang}}]{Dong2015}
{Dong}, R., {Hall}, C., {Rice}, K., \& {Chiang}, E. 2015, \apjl, 812, L32

\bibitem[{{Dong} {et~al.}(2022){Dong}, {Liu}, {Cuello}, {Pinte}, {{\'A}brah{\'a}m}, {Vorobyov}, {Hashimoto}, {K{\'o}sp{\'a}l}, {Chiang}, {Takami}, {Chen}, {Dunham}, {Fukagawa}, {Green}, {Hasegawa}, {Henning}, {Pavlyuchenkov}, {Pyo}, \& {Tamura}}]{Dong2022}
{Dong}, R., {Liu}, H.~B., {Cuello}, N., {et~al.} 2022, Nature Astronomy, 6, 331

\bibitem[{{Dullemond} {et~al.}(2019){Dullemond}, {K{\"u}ffmeier}, {Goicovic}, {Fukagawa}, {Oehl}, \& {Kramer}}]{Dullemond2019}
{Dullemond}, C.~P., {K{\"u}ffmeier}, M., {Goicovic}, F., {et~al.} 2019, \aap, 628, A20

\bibitem[{{Dunham} \& {Vorobyov}(2012)}]{Dunham2012}
{Dunham}, M.~M. \& {Vorobyov}, E.~I. 2012, \apj, 747, 52

\bibitem[{{Endres} {et~al.}(2016){Endres}, {Schlemmer}, {Schilke}, {Stutzki}, \& {M{\"u}ller}}]{cdms}
{Endres}, C.~P., {Schlemmer}, S., {Schilke}, P., {Stutzki}, J., \& {M{\"u}ller}, H. S.~P. 2016, Journal of Molecular Spectroscopy, 327, 95

\bibitem[{{Enoch} {et~al.}(2009){Enoch}, {Evans}, {Sargent}, \& {Glenn}}]{Enoch2009}
{Enoch}, M.~L., {Evans}, Neal~J., I., {Sargent}, A.~I., \& {Glenn}, J. 2009, \apj, 692, 973

\bibitem[{{Gahm} {et~al.}(2018){Gahm}, {Petrov}, {Tambovsteva}, {Grinin}, {Stempels}, \& {Walter}}]{Gahm2018}
{Gahm}, G.~F., {Petrov}, P.~P., {Tambovsteva}, L.~V., {et~al.} 2018, \aap, 614, A117

\bibitem[{{Gaia Collaboration} {et~al.}(2023){Gaia Collaboration}, {Vallenari}, {Brown}, {Prusti}, {de Bruijne}, {Arenou}, {Babusiaux}, {Biermann}, {Creevey}, {Ducourant}, {Evans}, {Eyer}, {Guerra}, {Hutton}, {Jordi}, {Klioner}, {Lammers}, {Lindegren}, {Luri}, {Mignard}, {Panem}, {Pourbaix}, {Randich}, {Sartoretti}, {Soubiran}, {Tanga}, {Walton}, {Bailer-Jones}, {Bastian}, {Drimmel}, {Jansen}, {Katz}, {Lattanzi}, {van Leeuwen}, {Bakker}, {Cacciari}, {Casta{\~n}eda}, {De Angeli}, {Fabricius}, {Fouesneau}, {Fr{\'e}mat}, {Galluccio}, {Guerrier}, {Heiter}, {Masana}, {Messineo}, {Mowlavi}, {Nicolas}, {Nienartowicz}, {Pailler}, {Panuzzo}, {Riclet}, {Roux}, {Seabroke}, {Sordo}, {Th{\'e}venin}, {Gracia-Abril}, {Portell}, {Teyssier}, {Altmann}, {Andrae}, {Audard}, {Bellas-Velidis}, {Benson}, {Berthier}, {Blomme}, {Burgess}, {Busonero}, {Busso}, {C{\'a}novas}, {Carry}, {Cellino}, {Cheek}, {Clementini}, {Damerdji}, {Davidson}, {de Teodoro}, {Nu{\~n}ez Campos}, {Delchambre}, {Dell'Oro}, {Esquej},
  {Fern{\'a}ndez-Hern{\'a}ndez}, {Fraile}, {Garabato}, {Garc{\'\i}a-Lario}, {Gosset}, {Haigron}, {Halbwachs}, {Hambly}, {Harrison}, {Hern{\'a}ndez}, {Hestroffer}, {Hodgkin}, {Holl}, {Jan{\ss}en}, {Jevardat de Fombelle}, {Jordan}, {Krone-Martins}, {Lanzafame}, {L{\"o}ffler}, {Marchal}, {Marrese}, {Moitinho}, {Muinonen}, {Osborne}, {Pancino}, {Pauwels}, {Recio-Blanco}, {Reyl{\'e}}, {Riello}, {Rimoldini}, {Roegiers}, {Rybizki}, {Sarro}, {Siopis}, {Smith}, {Sozzetti}, {Utrilla}, {van Leeuwen}, {Abbas}, {{\'A}brah{\'a}m}, {Abreu Aramburu}, {Aerts}, {Aguado}, {Ajaj}, {Aldea-Montero}, {Altavilla}, {{\'A}lvarez}, {Alves}, {Anders}, {Anderson}, {Anglada Varela}, {Antoja}, {Baines}, {Baker}, {Balaguer-N{\'u}{\~n}ez}, {Balbinot}, {Balog}, {Barache}, {Barbato}, {Barros}, {Barstow}, {Bartolom{\'e}}, {Bassilana}, {Bauchet}, {Becciani}, {Bellazzini}, {Berihuete}, {Bernet}, {Bertone}, {Bianchi}, {Binnenfeld}, {Blanco-Cuaresma}, {Blazere}, {Boch}, {Bombrun}, {Bossini}, {Bouquillon}, {Bragaglia}, {Bramante}, {Breedt},
  {Bressan}, {Brouillet}, {Brugaletta}, {Bucciarelli}, {Burlacu}, {Butkevich}, {Buzzi}, {Caffau}, {Cancelliere}, {Cantat-Gaudin}, {Carballo}, {Carlucci}, {Carnerero}, {Carrasco}, {Casamiquela}, {Castellani}, {Castro-Ginard}, {Chaoul}, {Charlot}, {Chemin}, {Chiaramida}, {Chiavassa}, {Chornay}, {Comoretto}, {Contursi}, {Cooper}, {Cornez}, {Cowell}, {Crifo}, {Cropper}, {Crosta}, {Crowley}, {Dafonte}, {Dapergolas}, {David}, {David}, {de Laverny}, {De Luise}, {De March}, {De Ridder}, {de Souza}, {de Torres}, {del Peloso}, {del Pozo}, {Delbo}, {Delgado}, {Delisle}, {Demouchy}, {Dharmawardena}, {Di Matteo}, {Diakite}, {Diener}, {Distefano}, {Dolding}, {Edvardsson}, {Enke}, {Fabre}, {Fabrizio}, {Faigler}, {Fedorets}, {Fernique}, {Fienga}, {Figueras}, {Fournier}, {Fouron}, {Fragkoudi}, {Gai}, {Garcia-Gutierrez}, {Garcia-Reinaldos}, {Garc{\'\i}a-Torres}, {Garofalo}, {Gavel}, {Gavras}, {Gerlach}, {Geyer}, {Giacobbe}, {Gilmore}, {Girona}, {Giuffrida}, {Gomel}, {Gomez}, {Gonz{\'a}lez-N{\'u}{\~n}ez},
  {Gonz{\'a}lez-Santamar{\'\i}a}, {Gonz{\'a}lez-Vidal}, {Granvik}, {Guillout}, {Guiraud}, {Guti{\'e}rrez-S{\'a}nchez}, {Guy}, {Hatzidimitriou}, {Hauser}, {Haywood}, {Helmer}, {Helmi}, {Sarmiento}, {Hidalgo}, {Hilger}, {H{\l}adczuk}, {Hobbs}, {Holland}, {Huckle}, {Jardine}, {Jasniewicz}, {Jean-Antoine Piccolo}, {Jim{\'e}nez-Arranz}, {Jorissen}, {Juaristi Campillo}, {Julbe}, {Karbevska}, {Kervella}, {Khanna}, {Kontizas}, {Kordopatis}, {Korn}, {K{\'o}sp{\'a}l}, {Kostrzewa-Rutkowska}, {Kruszy{\'n}ska}, {Kun}, {Laizeau}, {Lambert}, {Lanza}, {Lasne}, {Le Campion}, {Lebreton}, {Lebzelter}, {Leccia}, {Leclerc}, {Lecoeur-Taibi}, {Liao}, {Licata}, {Lindstr{\o}m}, {Lister}, {Livanou}, {Lobel}, {Lorca}, {Loup}, {Madrero Pardo}, {Magdaleno Romeo}, {Managau}, {Mann}, {Manteiga}, {Marchant}, {Marconi}, {Marcos}, {Marcos Santos}, {Mar{\'\i}n Pina}, {Marinoni}, {Marocco}, {Marshall}, {Martin Polo}, {Mart{\'\i}n-Fleitas}, {Marton}, {Mary}, {Masip}, {Massari}, {Mastrobuono-Battisti}, {Mazeh}, {McMillan}, {Messina}, {Michalik},
  {Millar}, {Mints}, {Molina}, {Molinaro}, {Moln{\'a}r}, {Monari}, {Mongui{\'o}}, {Montegriffo}, {Montero}, {Mor}, {Mora}, {Morbidelli}, {Morel}, {Morris}, {Muraveva}, {Murphy}, {Musella}, {Nagy}, {Noval}, {Oca{\~n}a}, {Ogden}, {Ordenovic}, {Osinde}, {Pagani}, {Pagano}, {Palaversa}, {Palicio}, {Pallas-Quintela}, {Panahi}, {Payne-Wardenaar}, {Pe{\~n}alosa Esteller}, {Penttil{\"a}}, {Pichon}, {Piersimoni}, {Pineau}, {Plachy}, {Plum}, {Poggio}, {Pr{\v{s}}a}, {Pulone}, {Racero}, {Ragaini}, {Rainer}, {Raiteri}, {Rambaux}, {Ramos}, {Ramos-Lerate}, {Re Fiorentin}, {Regibo}, {Richards}, {Rios Diaz}, {Ripepi}, {Riva}, {Rix}, {Rixon}, {Robichon}, {Robin}, {Robin}, {Roelens}, {Rogues}, {Rohrbasser}, {Romero-G{\'o}mez}, {Rowell}, {Royer}, {Ruz Mieres}, {Rybicki}, {Sadowski}, {S{\'a}ez N{\'u}{\~n}ez}, {Sagrist{\`a} Sell{\'e}s}, {Sahlmann}, {Salguero}, {Samaras}, {Sanchez Gimenez}, {Sanna}, {Santove{\~n}a}, {Sarasso}, {Schultheis}, {Sciacca}, {Segol}, {Segovia}, {S{\'e}gransan}, {Semeux}, {Shahaf}, {Siddiqui}, {Siebert},
  {Siltala}, {Silvelo}, {Slezak}, {Slezak}, {Smart}, {Snaith}, {Solano}, {Solitro}, {Souami}, {Souchay}, {Spagna}, {Spina}, {Spoto}, {Steele}, {Steidelm{\"u}ller}, {Stephenson}, {S{\"u}veges}, {Surdej}, {Szabados}, {Szegedi-Elek}, {Taris}, {Taylor}, {Teixeira}, {Tolomei}, {Tonello}, {Torra}, {Torra}, {Torralba Elipe}, {Trabucchi}, {Tsounis}, {Turon}, {Ulla}, {Unger}, {Vaillant}, {van Dillen}, {van Reeven}, {Vanel}, {Vecchiato}, {Viala}, {Vicente}, {Voutsinas}, {Weiler}, {Wevers}, {Wyrzykowski}, {Yoldas}, {Yvard}, {Zhao}, {Zorec}, {Zucker}, \& {Zwitter}}]{gaiadr3}
{Gaia Collaboration}, {Vallenari}, A., {Brown}, A.~G.~A., {et~al.} 2023, \aap, 674, A1

\bibitem[{{Galli} {et~al.}(2018){Galli}, {Loinard}, {Ortiz-L{\'e}on}, {Kounkel}, {Dzib}, {Mioduszewski}, {Rodr{\'\i}guez}, {Hartmann}, {Teixeira}, {Torres}, {Rivera}, {Boden}, {Evans}, {Brice{\~n}o}, {Tobin}, \& {Heyer}}]{Galli2018}
{Galli}, P. A.~B., {Loinard}, L., {Ortiz-L{\'e}on}, G.~N., {et~al.} 2018, \apj, 859, 33

\bibitem[{{Garufi} {et~al.}(2022){Garufi}, {Podio}, {Codella}, {Segura-Cox}, {Vander Donckt}, {Mercimek}, {Bacciotti}, {Fedele}, {Kasper}, {Pineda}, {Humphreys}, \& {Testi}}]{Garufi2022}
{Garufi}, A., {Podio}, L., {Codella}, C., {et~al.} 2022, \aap, 658, A104

\bibitem[{{Ginski} {et~al.}(2021){Ginski}, {Facchini}, {Huang}, {Benisty}, {Vaendel}, {Stapper}, {Dominik}, {Bae}, {M{\'e}nard}, {Muro-Arena}, {Hogerheijde}, {McClure}, {van Holstein}, {Birnstiel}, {Boehler}, {Bohn}, {Flock}, {Mamajek}, {Manara}, {Pinilla}, {Pinte}, \& {Ribas}}]{Ginski2021}
{Ginski}, C., {Facchini}, S., {Huang}, J., {et~al.} 2021, \apjl, 908, L25

\bibitem[{{Gupta} {et~al.}(2023){Gupta}, {Miotello}, {Manara}, {Williams}, {Facchini}, {Beccari}, {Birnstiel}, {Ginski}, {Hacar}, {K{\"u}ffmeier}, {Testi}, {Tychoniec}, \& {Yen}}]{Gupta2023}
{Gupta}, A., {Miotello}, A., {Manara}, C.~F., {et~al.} 2023, \aap, 670, L8

\bibitem[{{Hacar} {et~al.}(2023){Hacar}, {Clark}, {Heitsch}, {Kainulainen}, {Panopoulou}, {Seifried}, \& {Smith}}]{Hacar2023}
{Hacar}, A., {Clark}, S.~E., {Heitsch}, F., {et~al.} 2023, in Astronomical Society of the Pacific Conference Series, Vol. 534, Protostars and Planets VII, ed. S.~{Inutsuka}, Y.~{Aikawa}, T.~{Muto}, K.~{Tomida}, \& M.~{Tamura}, 153

\bibitem[{{Haugb{\o}lle} {et~al.}(2018){Haugb{\o}lle}, {Padoan}, \& {Nordlund}}]{Haugbolle2018}
{Haugb{\o}lle}, T., {Padoan}, P., \& {Nordlund}, {\r{A}}. 2018, \apj, 854, 35

\bibitem[{{Hennebelle} {et~al.}(2017){Hennebelle}, {Lesur}, \& {Fromang}}]{Hennebelle2017}
{Hennebelle}, P., {Lesur}, G., \& {Fromang}, S. 2017, \aap, 599, A86

\bibitem[{{Hsieh} {et~al.}(2023){Hsieh}, {Segura-Cox}, {Pineda}, {Caselli}, {Bouscasse}, {Neri}, {Lopez-Sepulcre}, {Valdivia-Mena}, {Maureira}, {Henning}, {Smirnov-Pinchukov}, {Semenov}, {M{\"o}ller}, {Cunningham}, {Fuente}, {Marino}, {Dutrey}, {Tafalla}, {Chapillon}, {Ceccarelli}, \& {Zhao}}]{Hsieh2023}
{Hsieh}, T.~H., {Segura-Cox}, D.~M., {Pineda}, J.~E., {et~al.} 2023, \aap, 669, A137

\bibitem[{{Huang} {et~al.}(2020){Huang}, {Andrews}, {{\"O}berg}, {Ansdell}, {Benisty}, {Carpenter}, {Isella}, {P{\'e}rez}, {Ricci}, {Williams}, {Wilner}, \& {Zhu}}]{Huang2020}
{Huang}, J., {Andrews}, S.~M., {{\"O}berg}, K.~I., {et~al.} 2020, \apj, 898, 140

\bibitem[{{Huang} {et~al.}(2023){Huang}, {Bergin}, {Bae}, {Benisty}, \& {Andrews}}]{Huang2023}
{Huang}, J., {Bergin}, E.~A., {Bae}, J., {Benisty}, M., \& {Andrews}, S.~M. 2023, \apj, 943, 107

\bibitem[{{Huang} {et~al.}(2021){Huang}, {Bergin}, {{\"O}berg}, {Andrews}, {Teague}, {Law}, {Kalas}, {Aikawa}, {Bae}, {Bergner}, {Booth}, {Bosman}, {Calahan}, {Cataldi}, {Cleeves}, {Czekala}, {Ilee}, {Le Gal}, {Guzm{\'a}n}, {Long}, {Loomis}, {M{\'e}nard}, {Nomura}, {Qi}, {Schwarz}, {Tsukagoshi}, {van't Hoff}, {Walsh}, {Wilner}, {Yamato}, \& {Zhang}}]{Huang2021}
{Huang}, J., {Bergin}, E.~A., {{\"O}berg}, K.~I., {et~al.} 2021, \apjs, 257, 19

\bibitem[{{Huang} {et~al.}(2022){Huang}, {Ginski}, {Benisty}, {Ren}, {Bohn}, {Choquet}, {{\"O}berg}, {Ribas}, {Bae}, {Bergin}, {Birnstiel}, {Boehler}, {Facchini}, {Harsono}, {Hogerheijde}, {Long}, {Manara}, {M{\'e}nard}, {Pinilla}, {Pinte}, {Rab}, {Williams}, \& {Zurlo}}]{Huang2022}
{Huang}, J., {Ginski}, C., {Benisty}, M., {et~al.} 2022, \apj, 930, 171

\bibitem[{{Jensen} \& {Haugb{\o}lle}(2018)}]{Jensen2018}
{Jensen}, S.~S. \& {Haugb{\o}lle}, T. 2018, \mnras, 474, 1176

\bibitem[{{J{\o}rgensen} {et~al.}(2004){J{\o}rgensen}, {Hogerheijde}, {Blake}, {van Dishoeck}, {Mundy}, \& {Sch{\"o}ier}}]{jorgensen04}
{J{\o}rgensen}, J.~K., {Hogerheijde}, M.~R., {Blake}, G.~A., {et~al.} 2004, \aap, 415, 1021

\bibitem[{{J{\o}rgensen} {et~al.}(2005){J{\o}rgensen}, {Sch{\"o}ier}, \& {van Dishoeck}}]{Jorgensen2005}
{J{\o}rgensen}, J.~K., {Sch{\"o}ier}, F.~L., \& {van Dishoeck}, E.~F. 2005, \aap, 435, 177

\bibitem[{{Kenyon} {et~al.}(1990){Kenyon}, {Hartmann}, {Strom}, \& {Strom}}]{Kenyon1990}
{Kenyon}, S.~J., {Hartmann}, L.~W., {Strom}, K.~M., \& {Strom}, S.~E. 1990, \aj, 99, 869

\bibitem[{{Kuffmeier} {et~al.}(2021){Kuffmeier}, {Dullemond}, {Reissl}, \& {Goicovic}}]{Kuffmeier2021}
{Kuffmeier}, M., {Dullemond}, C.~P., {Reissl}, S., \& {Goicovic}, F.~G. 2021, \aap, 656, A161

\bibitem[{{Kuffmeier} {et~al.}(2018){Kuffmeier}, {Frimann}, {Jensen}, \& {Haugb{\o}lle}}]{Kuffmeier2018}
{Kuffmeier}, M., {Frimann}, S., {Jensen}, S.~S., \& {Haugb{\o}lle}, T. 2018, \mnras, 475, 2642

\bibitem[{{Kuffmeier} {et~al.}(2017){Kuffmeier}, {Haugb{\o}lle}, \& {Nordlund}}]{Kuffmeier2017}
{Kuffmeier}, M., {Haugb{\o}lle}, T., \& {Nordlund}, {\r{A}}. 2017, \apj, 846, 7

\bibitem[{{Kuffmeier} {et~al.}(2023){Kuffmeier}, {Jensen}, \& {Haugb{\o}lle}}]{Kuffmeier2023}
{Kuffmeier}, M., {Jensen}, S.~S., \& {Haugb{\o}lle}, T. 2023, European Physical Journal Plus, 138, 272

\bibitem[{{Kuznetsova} {et~al.}(2022){Kuznetsova}, {Bae}, {Hartmann}, \& {Mac Low}}]{Kuznetsova2022}
{Kuznetsova}, A., {Bae}, J., {Hartmann}, L., \& {Mac Low}, M.-M. 2022, \apj, 928, 92

\bibitem[{{Kuznetsova} {et~al.}(2019){Kuznetsova}, {Hartmann}, \& {Heitsch}}]{Kuznetsova2019}
{Kuznetsova}, A., {Hartmann}, L., \& {Heitsch}, F. 2019, \apj, 876, 33

\bibitem[{{Lebreuilly} {et~al.}(2021){Lebreuilly}, {Hennebelle}, {Colman}, {Commer{\c{c}}on}, {Klessen}, {Maury}, {Molinari}, \& {Testi}}]{Lebreuilly2021}
{Lebreuilly}, U., {Hennebelle}, P., {Colman}, T., {et~al.} 2021, \apjl, 917, L10

\bibitem[{{Lee} {et~al.}(2023){Lee}, {Matsumoto}, {Kim}, {Lee}, {Harsono}, {Bae}, {Evans}, {Inutsuka}, {Choi}, {Tatematsu}, {Lee}, \& {Jaffe}}]{Lee2023}
{Lee}, J.-E., {Matsumoto}, T., {Kim}, H.-J., {et~al.} 2023, \apj, 953, 82

\bibitem[{{Manara} {et~al.}(2023){Manara}, {Ansdell}, {Rosotti}, {Hughes}, {Armitage}, {Lodato}, \& {Williams}}]{Manara2023}
{Manara}, C.~F., {Ansdell}, M., {Rosotti}, G.~P., {et~al.} 2023, in Astronomical Society of the Pacific Conference Series, Vol. 534, Protostars and Planets VII, ed. S.~{Inutsuka}, Y.~{Aikawa}, T.~{Muto}, K.~{Tomida}, \& M.~{Tamura}, 539

\bibitem[{{Manara} {et~al.}(2018){Manara}, {Morbidelli}, \& {Guillot}}]{Manara2018}
{Manara}, C.~F., {Morbidelli}, A., \& {Guillot}, T. 2018, \aap, 618, L3

\bibitem[{{Mendoza} {et~al.}(2009){Mendoza}, {Tejeda}, \& {Nagel}}]{Mendoza2009}
{Mendoza}, S., {Tejeda}, E., \& {Nagel}, E. 2009, \mnras, 393, 579

\bibitem[{{Mercimek} {et~al.}(2023){Mercimek}, {Podio}, {Codella}, {Chahine}, {L{\'o}pez-Sepulcre}, {Ohashi}, {Loinard}, {Johnstone}, {Menard}, {Cuello}, {Caselli}, {Zamponi}, {Aikawa}, {Bianchi}, {Busquet}, {Pineda}, {Bouvier}, {De Simone}, {Zhang}, {Sakai}, {Chandler}, {Ceccarelli}, {Alves}, {Dur{\'a}n}, {Fedele}, {Murillo}, {Jim{\'e}nez-Serra}, \& {Yamamoto}}]{Mercimek2023}
{Mercimek}, S., {Podio}, L., {Codella}, C., {et~al.} 2023, \mnras, 522, 2384

\bibitem[{{Mulders} {et~al.}(2021){Mulders}, {Pascucci}, {Ciesla}, \& {Fernandes}}]{Mulders2021}
{Mulders}, G.~D., {Pascucci}, I., {Ciesla}, F.~J., \& {Fernandes}, R.~B. 2021, \apj, 920, 66

\bibitem[{{Murillo} {et~al.}(2022){Murillo}, {van Dishoeck}, {Hacar}, {Harsono}, \& {J{\o}rgensen}}]{Murillo2022}
{Murillo}, N.~M., {van Dishoeck}, E.~F., {Hacar}, A., {Harsono}, D., \& {J{\o}rgensen}, J.~K. 2022, \aap, 658, A53

\bibitem[{{Nanne} {et~al.}(2019){Nanne}, {Nimmo}, {Cuzzi}, \& {Kleine}}]{Nanne2019N}
{Nanne}, J. A.~M., {Nimmo}, F., {Cuzzi}, J.~N., \& {Kleine}, T. 2019, Earth and Planetary Science Letters, 511, 44

\bibitem[{{Padoan} {et~al.}(2014){Padoan}, {Haugb{\o}lle}, \& {Nordlund}}]{Padoan2014}
{Padoan}, P., {Haugb{\o}lle}, T., \& {Nordlund}, {\r{A}}. 2014, \apj, 797, 32

\bibitem[{{Padoan} {et~al.}(2005){Padoan}, {Kritsuk}, {Norman}, \& {Nordlund}}]{Padoan2005}
{Padoan}, P., {Kritsuk}, A., {Norman}, M.~L., \& {Nordlund}, {\r{A}}. 2005, \apjl, 622, L61

\bibitem[{Pedregosa {et~al.}(2011)Pedregosa, Varoquaux, Gramfort, Michel, Thirion, Grisel, Blondel, Prettenhofer, Weiss, Dubourg, Vanderplas, Passos, Cournapeau, Brucher, Perrot, \& Duchesnay}]{scikit-learn}
Pedregosa, F., Varoquaux, G., Gramfort, A., {et~al.} 2011, Journal of Machine Learning Research, 12, 2825

\bibitem[{{Pelkonen} {et~al.}(2021){Pelkonen}, {Padoan}, {Haugb{\o}lle}, \& {Nordlund}}]{Pelkonen2021}
{Pelkonen}, V.~M., {Padoan}, P., {Haugb{\o}lle}, T., \& {Nordlund}, {\r{A}}. 2021, \mnras, 504, 1219

\bibitem[{{Pineda} {et~al.}(2023){Pineda}, {Arzoumanian}, {Andre}, {Friesen}, {Zavagno}, {Clarke}, {Inoue}, {Chen}, {Lee}, {Soler}, \& {Kuffmeier}}]{Pineda2023}
{Pineda}, J.~E., {Arzoumanian}, D., {Andre}, P., {et~al.} 2023, in Astronomical Society of the Pacific Conference Series, Vol. 534, Protostars and Planets VII, ed. S.~{Inutsuka}, Y.~{Aikawa}, T.~{Muto}, K.~{Tomida}, \& M.~{Tamura}, 233

\bibitem[{{Pineda} {et~al.}(2020){Pineda}, {Segura-Cox}, {Caselli}, {Cunningham}, {Zhao}, {Schmiedeke}, {Maureira}, \& {Neri}}]{Pineda2020}
{Pineda}, J.~E., {Segura-Cox}, D., {Caselli}, P., {et~al.} 2020, Nature Astronomy, 4, 1158

\bibitem[{{Rein} \& {Liu}(2012)}]{rebound}
{Rein}, H. \& {Liu}, S.~F. 2012, \aap, 537, A128

\bibitem[{{Sch{\"o}ier} {et~al.}(2005){Sch{\"o}ier}, {van der Tak}, {van Dishoeck}, \& {Black}}]{lamda}
{Sch{\"o}ier}, F.~L., {van der Tak}, F.~F.~S., {van Dishoeck}, E.~F., \& {Black}, J.~H. 2005, \aap, 432, 369

\bibitem[{{Seifried} {et~al.}(2013){Seifried}, {Banerjee}, {Pudritz}, \& {Klessen}}]{Seifried2013}
{Seifried}, D., {Banerjee}, R., {Pudritz}, R.~E., \& {Klessen}, R.~S. 2013, \mnras, 432, 3320

\bibitem[{{Shariff} {et~al.}(2022){Shariff}, {Gorti}, \& {Melon Fuksman}}]{Shariff2022}
{Shariff}, K., {Gorti}, U., \& {Melon Fuksman}, J.~D. 2022, \mnras, 514, 5548

\bibitem[{{Shu}(1977)}]{Shu1977}
{Shu}, F.~H. 1977, \apj, 214, 488

\bibitem[{{Tang} {et~al.}(2012){Tang}, {Guilloteau}, {Pi{\'e}tu}, {Dutrey}, {Ohashi}, \& {Ho}}]{Tang2012}
{Tang}, Y.~W., {Guilloteau}, S., {Pi{\'e}tu}, V., {et~al.} 2012, \aap, 547, A84

\bibitem[{{Terebey} {et~al.}(1984){Terebey}, {Shu}, \& {Cassen}}]{Terebey1984}
{Terebey}, S., {Shu}, F.~H., \& {Cassen}, P. 1984, \apj, 286, 529

\bibitem[{{Thieme} {et~al.}(2022){Thieme}, {Lai}, {Lin}, {Cheong}, {Lee}, {Yen}, {Li}, {Lam}, \& {Zhao}}]{Thieme2022}
{Thieme}, T.~J., {Lai}, S.-P., {Lin}, S.-J., {et~al.} 2022, \apj, 925, 32

\bibitem[{{Thies} {et~al.}(2011){Thies}, {Kroupa}, {Goodwin}, {Stamatellos}, \& {Whitworth}}]{Thies2011}
{Thies}, I., {Kroupa}, P., {Goodwin}, S.~P., {Stamatellos}, D., \& {Whitworth}, A.~P. 2011, \mnras, 417, 1817

\bibitem[{{Tobin} {et~al.}(2012){Tobin}, {Hartmann}, {Bergin}, {Chiang}, {Looney}, {Chandler}, {Maret}, \& {Heitsch}}]{Tobin2012}
{Tobin}, J.~J., {Hartmann}, L., {Bergin}, E., {et~al.} 2012, \apj, 748, 16

\bibitem[{{Tokuda} {et~al.}(2018){Tokuda}, {Onishi}, {Saigo}, {Matsumoto}, {Inoue}, {Inutsuka}, {Fukui}, {Machida}, {Tomida}, {Hosokawa}, {Kawamura}, \& {Tachihara}}]{Tokuda2018}
{Tokuda}, K., {Onishi}, T., {Saigo}, K., {et~al.} 2018, \apj, 862, 8

\bibitem[{{Ulrich}(1976)}]{Ulrich1976}
{Ulrich}, R.~K. 1976, \apj, 210, 377

\bibitem[{{Unno} {et~al.}(2022){Unno}, {Hanawa}, \& {Takasao}}]{Unno2022}
{Unno}, M., {Hanawa}, T., \& {Takasao}, S. 2022, \apj, 941, 154

\bibitem[{{Valdivia-Mena} {et~al.}(2022){Valdivia-Mena}, {Pineda}, {Segura-Cox}, {Caselli}, {Neri}, {L{\'o}pez-Sepulcre}, {Cunningham}, {Bouscasse}, {Semenov}, {Henning}, {Pi{\'e}tu}, {Chapillon}, {Dutrey}, {Fuente}, {Guilloteau}, {Hsieh}, {Jim{\'e}nez-Serra}, {Marino}, {Maureira}, {Smirnov-Pinchukov}, {Tafalla}, \& {Zhao}}]{Valdivia-Mena2022}
{Valdivia-Mena}, M.~T., {Pineda}, J.~E., {Segura-Cox}, D.~M., {et~al.} 2022, \aap, 667, A12

\bibitem[{Virtanen {et~al.}(2020)Virtanen, Gommers, Oliphant, Haberland, Reddy, Cournapeau, Burovski, Peterson, Weckesser, Bright, {van der Walt}, Brett, Wilson, Millman, Mayorov, Nelson, Jones, Kern, Larson, Carey, Polat, Feng, Moore, {VanderPlas}, Laxalde, Perktold, Cimrman, Henriksen, Quintero, Harris, Archibald, Ribeiro, Pedregosa, {van Mulbregt}, \& {SciPy 1.0 Contributors}}]{scipy}
Virtanen, P., Gommers, R., Oliphant, T.~E., {et~al.} 2020, Nature Methods, 17, 261

\bibitem[{{Visser} {et~al.}(2009){Visser}, {van Dishoeck}, {Doty}, \& {Dullemond}}]{Visser2009}
{Visser}, R., {van Dishoeck}, E.~F., {Doty}, S.~D., \& {Dullemond}, C.~P. 2009, \aap, 495, 881

\bibitem[{{Vorobyov} \& {Basu}(2005)}]{Vorobyov2005}
{Vorobyov}, E.~I. \& {Basu}, S. 2005, \apjl, 633, L137

\bibitem[{{Vorobyov} {et~al.}(2020){Vorobyov}, {Skliarevskii}, {Elbakyan}, {Takami}, {Liu}, {Liu}, \& {Akiyama}}]{Vorobyov2020}
{Vorobyov}, E.~I., {Skliarevskii}, A.~M., {Elbakyan}, V.~G., {et~al.} 2020, \aap, 635, A196

\bibitem[{{Welch} {et~al.}(2000){Welch}, {Hartmann}, {Helfer}, \& {Brice{\~n}o}}]{Welch2000}
{Welch}, W.~J., {Hartmann}, L., {Helfer}, T., \& {Brice{\~n}o}, C. 2000, \apj, 540, 362

\bibitem[{{Yen} {et~al.}(2019){Yen}, {Gu}, {Hirano}, {Koch}, {Lee}, {Liu}, \& {Takakuwa}}]{Yen2019}
{Yen}, H.-W., {Gu}, P.-G., {Hirano}, N., {et~al.} 2019, \apj, 880, 69

\bibitem[{{Yen} {et~al.}(2018){Yen}, {Koch}, {Manara}, {Miotello}, \& {Testi}}]{Yen2018}
{Yen}, H.-W., {Koch}, P.~M., {Manara}, C.~F., {Miotello}, A., \& {Testi}, L. 2018, \aap, 616, A100

\bibitem[{{Yen} {et~al.}(2017){Yen}, {Takakuwa}, {Chu}, {Hirano}, {Ho}, {Kanagawa}, {Lee}, {Liu}, {Liu}, {Matsumoto}, {Matsushita}, {Muto}, {Saigo}, {Tang}, {Trejo}, \& {Wu}}]{yen2017}
{Yen}, H.-W., {Takakuwa}, S., {Chu}, Y.-H., {et~al.} 2017, \aap, 608, A134

\bibitem[{{Yen} {et~al.}(2014){Yen}, {Takakuwa}, {Ohashi}, {Aikawa}, {Aso}, {Koyamatsu}, {Machida}, {Saigo}, {Saito}, {Tomida}, \& {Tomisaka}}]{Yen2014}
{Yen}, H.-W., {Takakuwa}, S., {Ohashi}, N., {et~al.} 2014, \apj, 793, 1

\bibitem[{{Zhang} {et~al.}(2023){Zhang}, {Ginski}, {Huang}, {Zurlo}, {Beust}, {Bae}, {Benisty}, {Garufi}, {Hogerheijde}, {van Holstein}, {Kenworthy}, {Langlois}, {Manara}, {Pinilla}, {Rab}, {Ribas}, {Rosotti}, \& {Williams}}]{Zhang2023}
{Zhang}, Y., {Ginski}, C., {Huang}, J., {et~al.} 2023, \aap, 672, A145

\end{thebibliography}

\begin{appendix}

\section{Mendoza equations} \label{app:equations}

We use the equations derived in \citet{Mendoza2009} to compute infalling trajectories, as discussed in Sect. \ref{sec:model}. The equations were expressed in spherical coordinates $r$, $\theta,$ and $\phi$, which represent the radial coordinate, the polar angle, and the azimuthal angle, respectively (also see Fig. \ref{fig:coordinates}). The initial position of the infalling particle is then given as $r_0$, $\theta_0$, and $\phi_0$. The source of gravity (protostar) is set to be at the origin.

To begin with, \citet{Mendoza2009} defined two dimensionless parameters, \( \mu \) and \( \nu, \)
as
\begin{equation}
  \mu^2 \equiv \frac{h_0^2 }{ r_0^2  E_0} = \frac{ r_\text{u}^2 }{ r_0^2 }, 
  \qquad \nu^2 \equiv \frac{ v_{r_0}^2 }{ E_0 }, 
\label{eq03}
\end{equation}
where, $h_0$ is the initial specific angular momentum w.r.t. azimuthal axis (z-axis in Fig. \ref{fig:coordinates}). \( r_\text{u} \equiv h_0^2/GM \), can be thought of as the disk's radius in the UCM model and \(E_0 \equiv GM / r_\text{u} \) is the specific gravitational potential energy of infalling gas at \( r_\text{u} \).
In the following equations, distances are measured in the units of \( r_\text{u} \) and velocities are measured in the units of \(\sqrt{E_0} \) (Keplerian velocity at \(r_\text{u} \)).

% With these parameters, they expressed a dimensionless energy \(
% \varepsilon \equiv 2 E / E_0 \) as:
% \begin{equation}
%   \varepsilon = \nu^2 + \mu^2 \sin^2 \theta_0 - 2 \mu,
% \label{eq04}
% \end{equation}

% Assuming, initially the particle is located at \( r_0 \) distance from the source of gravity and its polar and azimuthal angles are \( \theta_0 \) and \( \phi_0 \), respectively. 
Over the course of particle's motion, the trajectory was defined as a function of the parametric azimuthal angle, \( \varphi \).
The trajectory of an infalling particle, given by equations of conic sections, was then represented as\begin{equation}
  r = \frac{ \sin^2 \theta_0 }{ 1 - e \cos\varphi },
\label{eq05}
\end{equation}
with the eccentricity, \( e, \) of the orbit given by
\begin{equation}
  e = \sqrt{ 1 + \varepsilon \sin^2 \theta_0 }.
\label{eq06}
\end{equation}
Here, $\varepsilon$ represents a dimensionless energy parameter, calculated as $\varepsilon = \nu^2 + \mu^2 \sin^2 \theta_0 - 2 \mu$.

\noindent At the border of the cloud, \( r = r_0 = 1/\mu \). After substituting this in Eq.~\eqref{eq05} and performing some spatial rotations, the following formulae
were obtained:
\begin{equation}
\cos( \varphi - \varphi_0 ) = \frac{ \cos\theta }{ \cos\theta_0 }
,\qquad \cos( \phi - \phi_0 ) = \frac{ \tan\theta_0 }{ \tan\theta }
\label{eq08}
.\end{equation}

% Using equation~\eqref{eq08}, equation~\eqref{eq05} was rewritten as:
% \begin{equation}
%   r = \frac{ \sin^2 \theta_0 }{ 1 - e \cos\xi },
% \label{eq09}
% \end{equation}
% \noindent where 
% \begin{equation}
%   \xi = \cos^{-1} \left( \frac{ \cos\theta }{ \cos\theta_0 }
% \right) + \varphi_0.
% \label{eq10}
% \end{equation}

Using the previous equations and standard definitions of azimuthal ($v_\phi$), polar ($v_\theta$), and radial ($v_r$) components of a velocity vector, the equations for velocities were derived as\begin{equation}
v_\phi = \frac{ \sin^2 \theta_0 }{ r \sin\theta },
\label{eq11}
\end{equation}
\begin{equation}
v_\theta = \frac{ \sin \theta_0 }{ r \sin \theta }
\left( \cos^2 \theta_0 - \cos^2 \theta \right)^{ 1/2 },
\label{eq12}
\end{equation}
\begin{equation}
v_r = -\frac{ e \sin \xi \sin \theta_0 }{ r ( 1 - e \cos\xi ) },
\label{eq13}
\end{equation}
\noindent where 
\begin{equation}
  \xi = \cos^{-1} \left( \frac{ \cos\theta }{ \cos\theta_0 }
\right) + \varphi_0.
\label{eq10}
\end{equation}We used these equations to compute the positions (Eqs. \ref{eq05} to \ref{eq08}) and velocities (Eqs. \ref{eq11} to \ref{eq13}) of infalling particles.

\section{3D morphology} \label{app:rebound}

\begin{figure*} 
\centering
\begin{subfigure}[b]{0.49\textwidth}
\centering
    \includegraphics[width=1.1\textwidth]{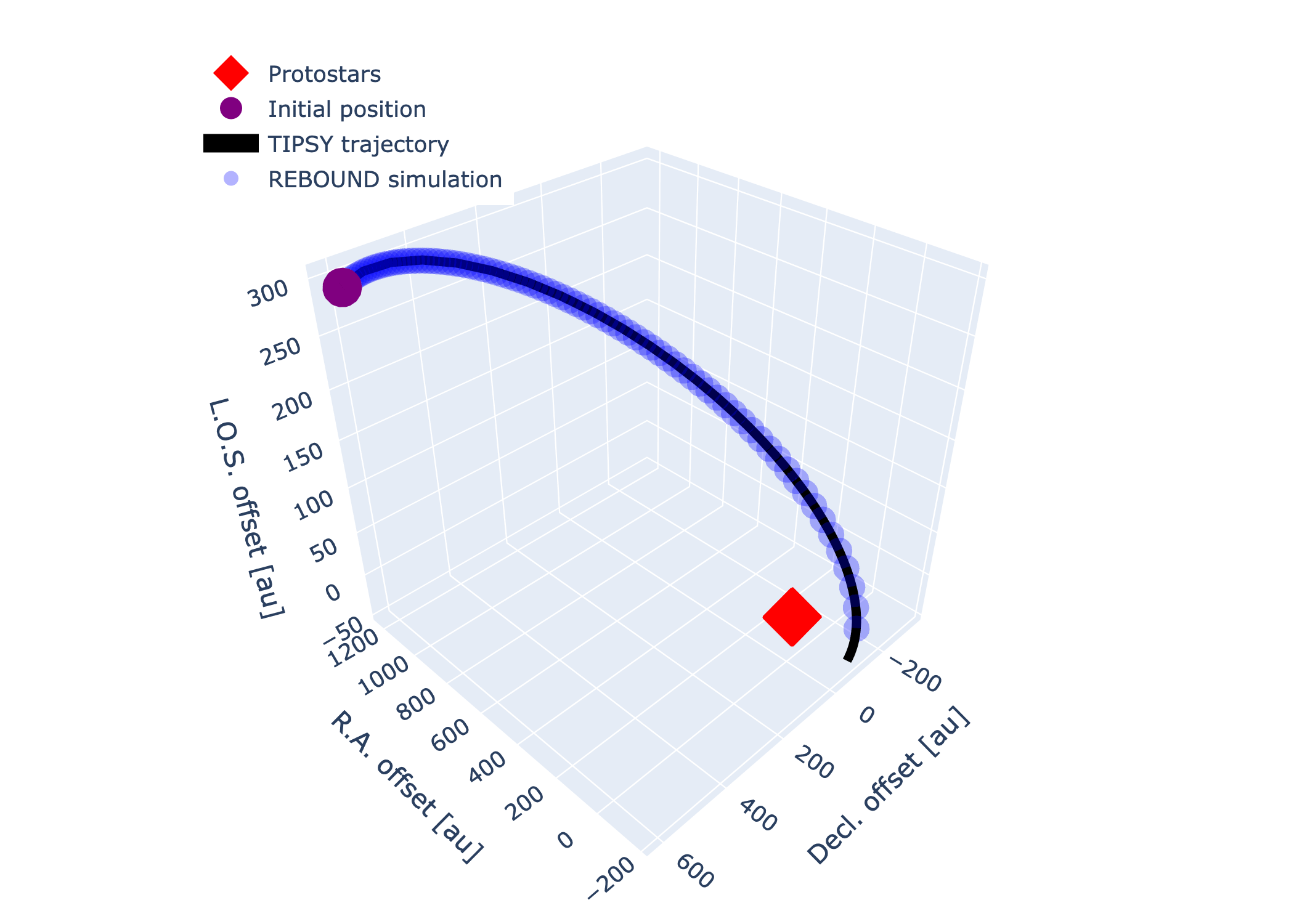}
    \caption{}
    \label{fig:SCrA_rebound}
    \end{subfigure}
% \hfill
\begin{subfigure}[b]{0.49\textwidth}
\centering
    \includegraphics[width=1.1\textwidth]{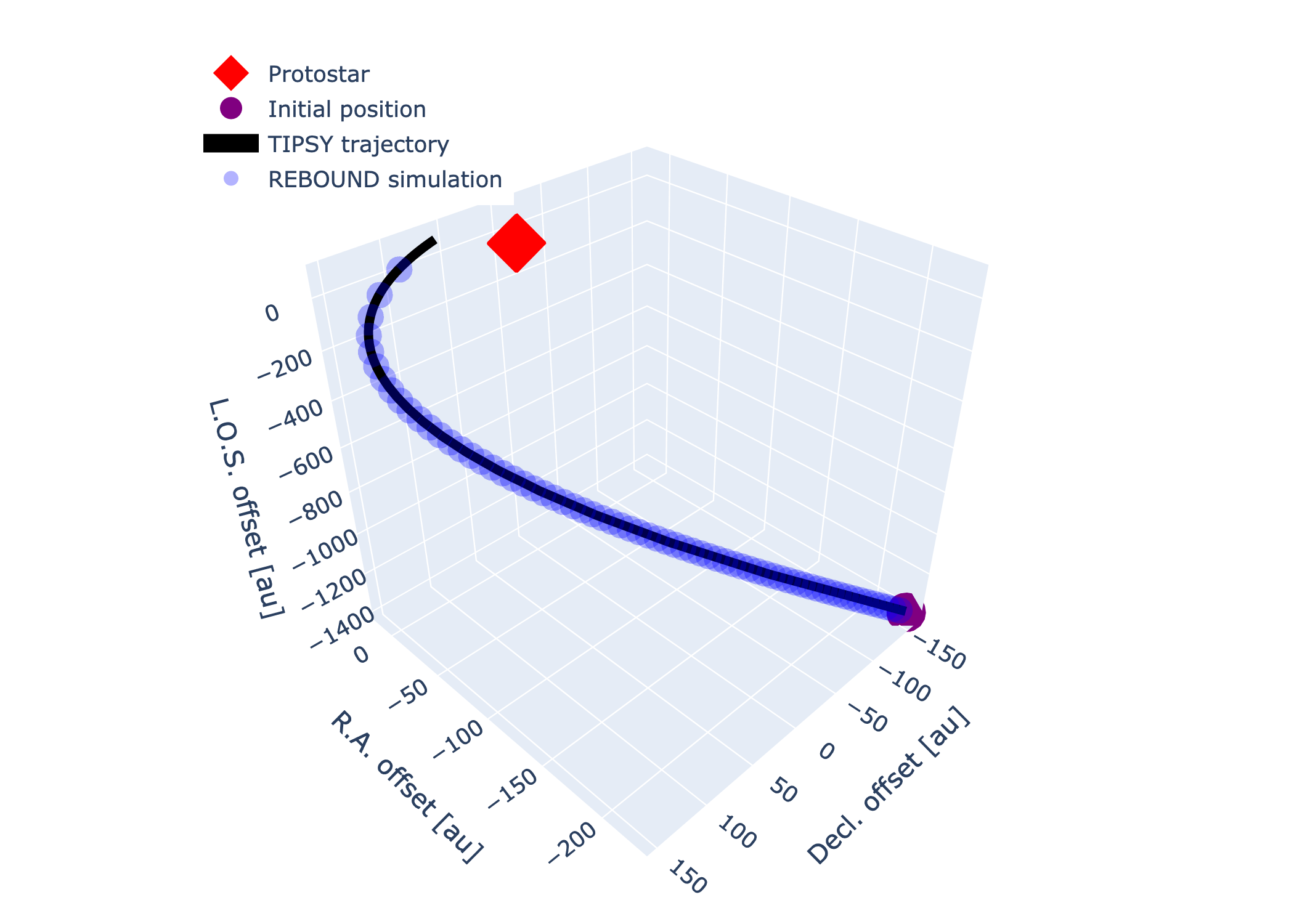}
    \caption{}
    \label{fig:hltau_rebound}
    \end{subfigure}
\caption{
Isometric projection of the best-fit infalling trajectory for streamers around S CrA (left panel) and HL Tau (right panel), in 3D position--position--position space (RA, Decl., and LOS or radial distance). The black line represents the analytical trajectory from our implementation of the \citet{Mendoza2009} models, as described in Sect. \ref{sec:model}. Blue spheres represent solutions from two-body REBOUND simulations \citep{rebound}. The red diamonds denote the position of the center of mass of the protostellar systems. The purple circles denote the initial position of the infalling gas. These trajectories are computed up to the closest approach of infalling material to the protostellar system. 
}
\label{fig:rebound}
\end{figure*}

The best-fit trajectories from TIPSY can also be used to infer the trajectory of infalling gas in 3D position--position--position space (RA, Decl., and LOS distance), as shown in Fig. \ref{fig:rebound}. As we expect all of the observed gas in these streamers to have similar initial conditions, these 3D trajectories represent the 3D morphologies of infalling streamers. 

Figure  \ref{fig:rebound}  also compares the 3D trajectory as inferred from our implementation of \citet{Mendoza2009} models (see Sect. \ref{sec:model}) to the solutions for the same initial configuration from simple two-body REBOUND simulations \citep{rebound}. Both the solutions are always in good agreement, suggesting that our implementation of \citet{Mendoza2009} models gives an accurate description of infalling particle motion. We note that the REBOUND simulations generally take $\gtrsim100$ times more time to compute solutions, making its use much less feasible for fitting streamers.

% ...mention rebound parameters...
% \section{S CrA $^{12}$CO map} \label{app:rebound}

\section{Distance metric} \label{app:dist}

\begin{figure*} 
\centering
\begin{subfigure}[b]{1.1\textwidth}
    \centering
    \includegraphics[width=\textwidth]{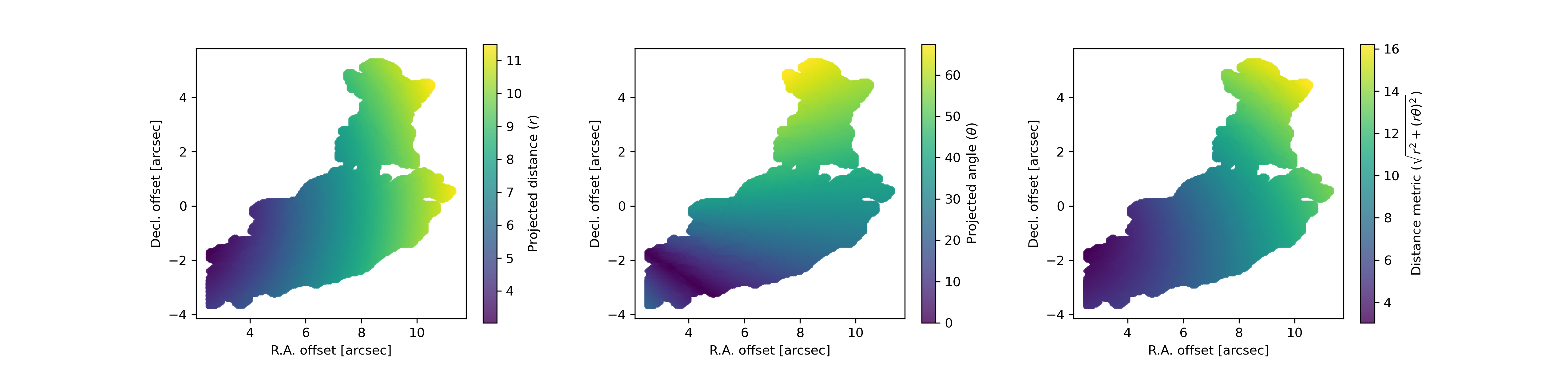}
    \caption{}
    \label{fig:SCrA_dist}
    \end{subfigure}
% \hfill
\begin{subfigure}[b]{1.1\textwidth}
    \centering
    \includegraphics[width=\textwidth]{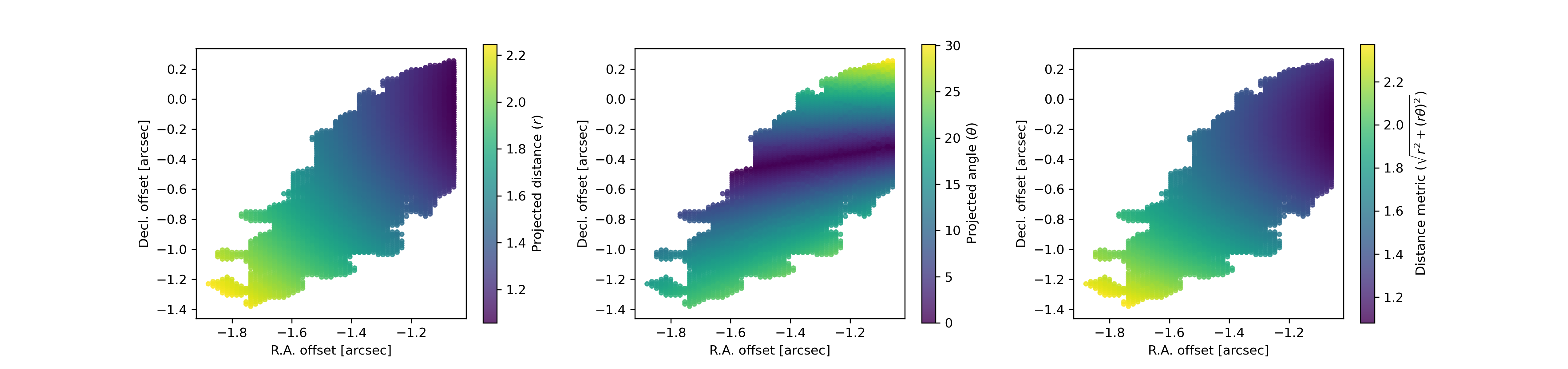}
    \caption{}
    \label{fig:hltau_dist}
    \end{subfigure}
\caption{Computation of the distance metric ($d$) using polar coordinates $r$ and $\theta$, as discussed in Appendix \ref{app:dist}, for S CrA (top panels) and HL Tau (bottom panels). 
\textit{Left panels}: Radial distance ($r = \sqrt{(\Delta RA)^{2} +(\Delta Decl.)^{2}}$) for each point on a streamer from the origin point (0,0). The origin point is also assumed to be the position of the center of mass for the protostellar system. 
\textit{Middle panels}: Polar angle ($\theta = \arctan{(\Delta Decl./\Delta RA)}$) for each point on a streamer with respect to the mean direction of the streamer points in the bin closest to the origin.
\textit{Right panels}: Distance metric ($d=\sqrt{r^{2}+(wr\theta)^{2}}$) for each point on a streamer computed using the polar coordinates ($r$ and $\theta$) and the weighting factor ($w$, $=1$ by default).
}
\label{fig:dist}
\end{figure*}

Figure \ref{fig:dist} illustrates computation of distance metric ($d=\sqrt{r^{2}+(wr\theta)^{2}}$), where $r$ and $\theta$ denote the projected radial distance (from the protostar) and polar angle (with respect to the median orientation of streamer points closer than the 10$^{th}$ percentile of $r$ distribution), respectively.
% polar coordinates of the points on streamer, as projected on the POS. 
The weighting factor ($w$, $=1$ by default) sets the importance of $r\theta$ (distance in azimuthal direction) in the computation of distance metric. Setting $w=0$, will set distance metric to be equal to the projected radial distance ($r$), similar to the approach by \citet{Yen2019}. 

Overall, a higher-value distance metric should correspond to the part of the streamer that is expected to be physically farthest away from the protostar(s). As discussed in Sect. \ref{sec:code}, this distance metric is used to bin the data (for computing intensity-weighted means and standard deviations) and as an independent parameter for the final fitting.
% \FloatBarrier
% \begin{multicols}{1}
% \begin{strip} 

\section{Stellar parameters for S CrA and HL Tau} \label{app:stellar_params}

Stellar parameters used for fitting the streamers are fixed before running TIPSY, as listed for S CrA and HL Tau in Table \ref{tab:params}. Stellar mass estimates for S CrA and HL Tau were taken from \citet{Gahm2018} and \citet{Yen2019}, respectively. Distance estimate for S CrA is based on \textit{Gaia} DR3 parallax value \citep{gaiadr3}. \textit{Gaia} measurements were unavailable for HL Tau, so we used the estimate of the distance to its surrounding cloud, Lynds 155 \citep{Galli2018}. 

% Systemic radial velocities for the S CrA A (northern protostar) and S CrA B (southern protostar) were inferred to be $6.07\pm0.09$~km~s~$^{-1}$ and $5.66 \pm 0.14$~km~s~$^{-1}$, respectively. These velocities were taken as the peak of Gaussian functions, by fitting them to C$^{18}$O (2--1) spectra of disks. These  C$^{18}$O (2--1) observations were taken as part of the same ALMA program, as the $^{13}$CO (2--1) observations discussed in Sec. \ref{sec:scra}.

Systemic LOS velocities for S CrA A (the northern protostar) and S CrA B (the southern protostar) were inferred to be $6.07\pm0.09$~km~s~$^{-1}$ and $5.66 \pm 0.14$~km~s~$^{-1}$, respectively, using the peak of Gaussian fits to C$^{18}$O (2--1) disk spectra. These C$^{18}$O (2--1) observations were part of the same ALMA project (Project Id.: 2019.1.01792.S) as the $^{13}$CO (2--1) observations discussed in Sect. \ref{sec:scra}. We used the mean systemic velocity of 5.86~km~s~$^{-1}$ for the TIPSY fitting of S CrA streamer. For HL Tau, the systemic velocity derived by \citet{Yen2019} was used.

% The parameters used to select the subcube with streamer emission, i.e., RA, Decl., and RV. limits, were decided using visual inspection of streamer emission in the PPV diagram (e.g., Fig. \ref{fig:tipsy_scra} \& \ref{fig:tipsy_hltau}). 
% Finally, a significance ($\sigma$) level (generally $\gtrsim3$) is used to only consider the pixels with a significant emission.

% After selecting the subcube, TIPSY further selects the emission above a certain noise ($\sigma$) level, as the emission of the streamer.

% Table \ref{tab:params} lists the fixed parameters used for fitting the streamers around S CrA and HL Tau, as discussed in Sec. \ref{sec:methodology}. Stellar mass estimates for S CrA and HL Tau were take from \citet{Gahm2018} and \citet{Yen2019}, respectively. Distance estimate for S CrA is based in \textit{Gaia} DR3 values \citep{gaiadr3}. \textit{Gaia} measurements were unavailable for HL Tau, so we use the distance estimate to it's surrounding cloud Lynds 155 \citep{Galli2018}. Systemic radial velocities for both of the sources were inferred from the peak of the spectra, focused on the disks.

% The parameters used to select the subcube of disk emission, i.e., RA, Decl., and RV limits of the streamer emission, are also listed in Table \ref{tab:params}. After selecting the subcube, TIPSY further selects the emission above a certain noise ($\sigma$) level, as the emission of the streamer.

\section{Integrated intensity maps} \label{app:moment0}

\begin{figure*} 
\centering
\begin{subfigure}[b]{0.49\textwidth}
\centering
    \includegraphics[width=1\textwidth]{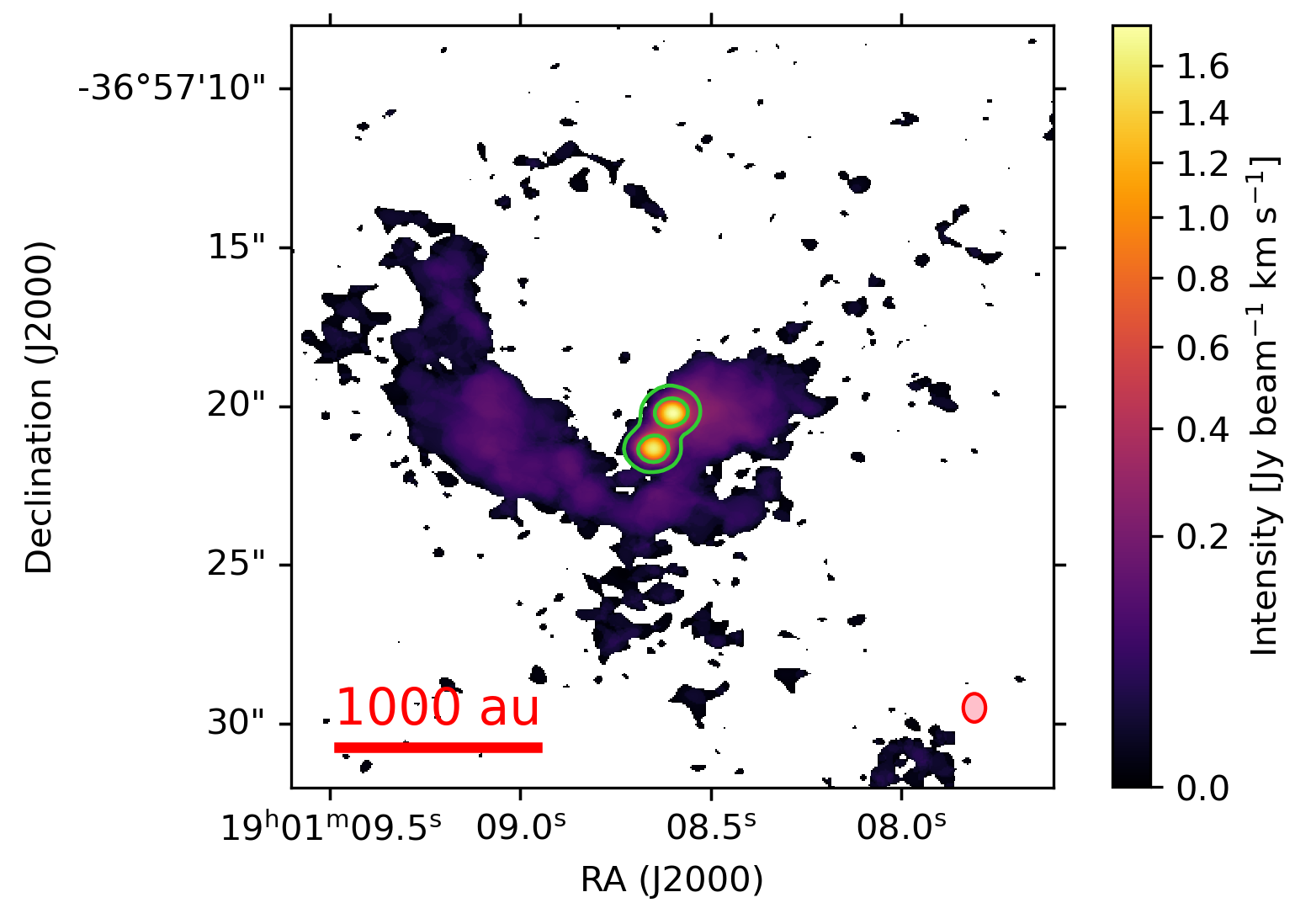}
    \caption{}
    \label{fig:SCrA_m0}
    \end{subfigure}
% \hfill
\begin{subfigure}[b]{0.49\textwidth}
\centering
    \includegraphics[width=1\textwidth]{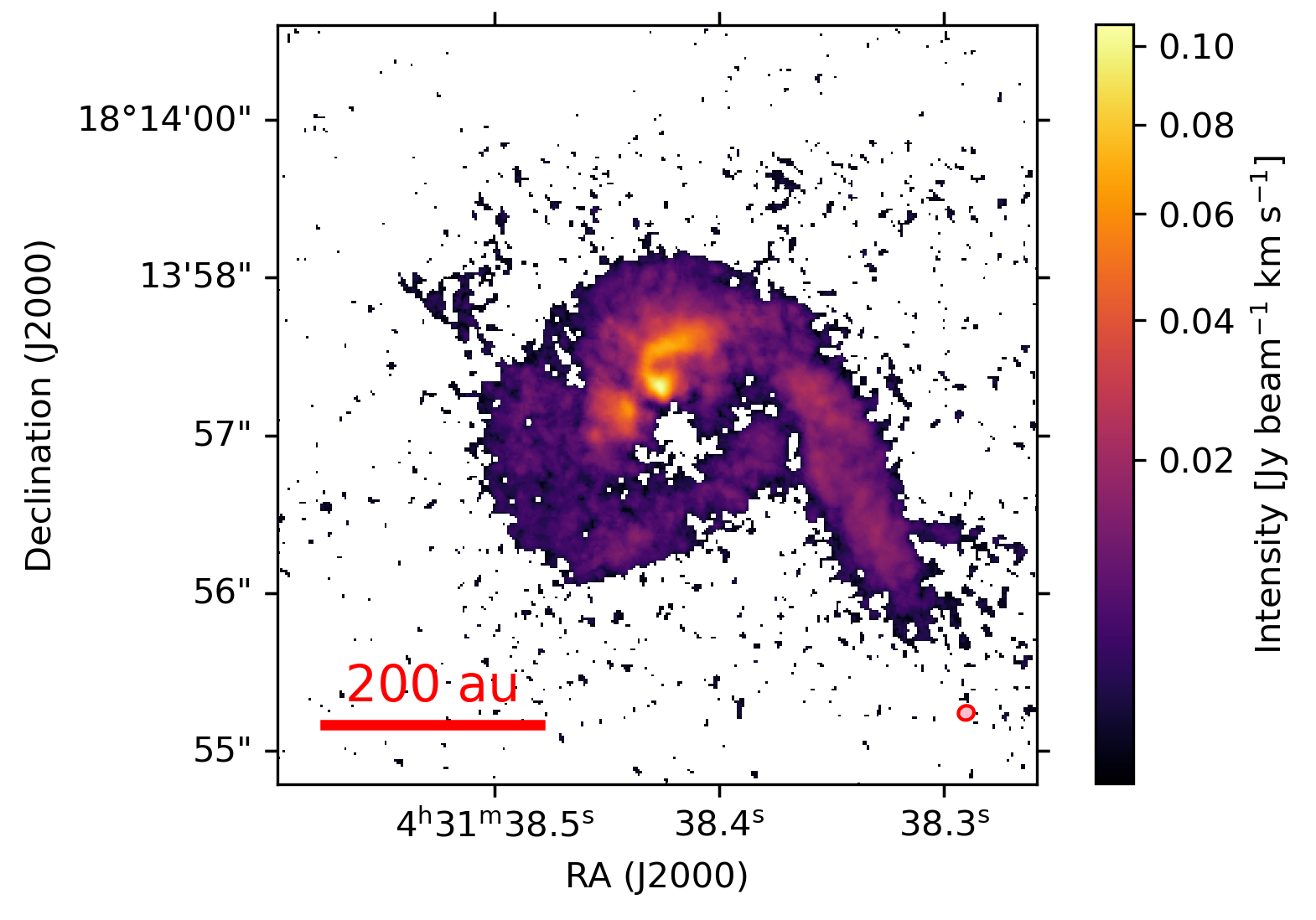}
    \caption{}
    \label{fig:hltau_m0}
    \end{subfigure}
\caption{
Integrated intensity (moment 0) maps for S CrA (left panel) and HL Tau (right panel), considering only pixels with an intensity $>3.5\sigma$. 
The horizontal red lines in the bottom-left corners represent the physical length scales, and the pink ellipses in the bottom-right corners represent the beam size.
Green contours in the left panel denote the continuum emission from the protoplanetary disks.
}
\label{fig:moment0}
\end{figure*}

Figure \ref{fig:moment0} shows integrated intensity (moment 0) maps for $13$CO (2--1) observations of S CrA and HCO$^{+}$ (3--2) observations of HL Tau. The streamers are visible as the elongated gas structures.

\end{appendix}

\end{document}